\documentclass[aps,a4paper,english, twocolumn,prb, showpacs, 10pt ]{revtex4}

\usepackage[latin1]{inputenc}
\usepackage{graphicx}
\usepackage{amsmath,amsfonts,amssymb,dsfont,amsxtra,amsopn,amstext,amsbsy}
\usepackage[english]{babel}
\usepackage{bm}
\usepackage{subfigure}
\usepackage{hyperref}
\usepackage{color}

\newcommand{\ua}{\uparrow}
\newcommand{\da}{\downarrow}
\newcommand{\ew}[1]{\left\langle #1 \right\rangle}

\newcommand{\dint}{\text{d}}
\newcommand{\im}{\text{Im}}

\newcommand{\ndg}{{\phantom{\dagger}}}
\newcommand{\dg}{\dagger}
\newcommand{\bqn}{\begin{eqnarray}}
\newcommand{\eqn}{\end{eqnarray}}
\newcommand{\bee}{\begin{equation}}
\newcommand{\ene}{\end{equation}}
\newcommand{\bal}{\begin{align}}
\newcommand{\eal}{\end{align}}
\newcommand{\ket}[1]{\left| #1 \right\rangle}
\newcommand{\bra}[1]{\left\langle #1 \right|}
\newcommand{\erzqd}[2]{\hat #1^\dagger _{#2}}
\newcommand{\verqd}[2]{\hat #1^\ndg _{#2}}

\newcommand{\hh}{\left|H\right\rangle}
\newcommand{\vv}{\left|V\right\rangle}
\newcommand{\qq}{\left|Q\right\rangle}
\newcommand{\bb}{\left|B\right\rangle}

\newcommand{\xii}{\left|X_i\right\rangle}
\newcommand{\xh}{\left|X_H\right\rangle}
\newcommand{\xv}{\left|X_V\right\rangle}
\newcommand{\GG}{\left|G\right\rangle}
\newcommand{\nt}{\notag}
\newcommand{\phdg}{{\phantom{\dagger}}}
\newcommand{\vex}{V_\text{ex}}

\renewcommand{\vec}[1]{\bm{\mathbf{#1}}}

\newcommand{\ii}{\mbox{\upshape i}}

\begin{document}

\title{Formation dynamics of an entangled photon pair -- a temperature dependent analysis}
\date{\today}
\author{A. Carmele}
\email [E-mail at: ]{alex@itp.physik.tu-berlin.de}
\author{F. Milde}
\author{M.-R. Dachner}
\affiliation{Institut f\"ur Theoretische Physik, Nichtlineare
Optik und Quantenelektronik,\\
Technische Universit\"at Berlin, Hardenbergstra{\ss}e 36,
EW 7-1 10623 Berlin, Germany}
\author{M. Bagheri Harouni}
\author{R. Rokniknizadeh}
\affiliation{Physics Department, Quantum Optics Group\\
University of Isfahan, 81746 Isfahan, Iran}
\author{M. Richter}
\author{A. Knorr}
\affiliation{Institut f\"ur Theoretische Physik, Nichtlineare
Optik und Quantenelektronik,\\
Technische Universit\"at Berlin, Hardenbergstra{\ss}e 36,
EW 7-1 10623 Berlin, Germany}

\begin{abstract}
We theoretically study the polarization entanglement of photons generated by the biexciton cascade in a GaAs/InAs semiconductor quantum dot (QD), located in a nano cavity. 
A detailed analysis of the complex interplay between photon- and carrier coherences and phonons which occurs during the cascade allows us to clearly identify where the entanglement is generated and destroyed.
A quantum state tomography is performed for varying exciton fine structure splittings.
By constructing an effective multi-phonon Hamiltonian which couples the continuum of the wetting layer states to the QD we investigate the relaxation of the biexciton and exciton states. This consistently introduces a temperature dependence to the cascade. Considering typical Stranski-Karastanov grown QDs, for temperatures around 80~K the degree of entanglement starts to be affected by the dephasing of the exciton states and is ultimately lost above 120~K.
\end{abstract}

\pacs{78.67.Hc, 42.50.Dv, 63.22.-m, 71.35.-y}
\maketitle
\section{Introduction}
Coherent superpositions of quantum states allow to outperform the classical bit. Based on this feature the quantum bit (qubit) was introduced.\cite{Nielsen::00} In light of integrated quantum communication, natural candidates for its realization seem to be photonic systems. Not only do their two different polarizations serve as basis states for qubits in quantum computation, but also has encoding and manipulating of quantum information on photons made tremendous advancement: In recent years, photons as qubits have been sent successfully over fiber communication channels,\cite{Pittman:PhysRevA:02,Stucki:NewJPhys:02} quantum cryptography is already technically feasible \cite{Gisin:RevModPhys:02,Poppe:OptExpress:04} and quantum teleportation of so-called entangled photon states was demonstrated.\cite{Bouwmeester:Nature:97,Bennett:PhysRevLett:93} Supported is this development by the convenient fact that linear quantum optics is sufficient, when implementing quantum computing algorithms.\cite{Scheel:PhysRevA:06}

Entanglement in its simplest form is a non-separable superposition of joint quantum states, in our case qubits, that show non-local quantum correlation. Among different proposals \cite{Brendel:PhysRevLett:99,Edamatsu:Nature:04,Fasel:NewJPhys:04,Ou:PhysRevLett:88,Shih:PhysRevLett:88} very promising solid state source for entangled photon pairs are semiconductor quantum dots (QDs).\cite{Benson:PhysRevLett:00}  In contrast to parametric down conversion,\cite{Burnham:PhysRevLett:70} where entangled pairs are produced in a probabilistic manner with low efficiency ($10^{-10}$ parametric photons per pump photon\cite{Bayer:PhysRevLett:01}) radiative decay of a biexciton cascade in QDs provides an on-demand generation of entanglement, a crucial requirement for scalable quantum networks.

The prospect of all-integrated photonic applications in combination with compact semiconductor devices raises the question of how robust and efficient an embedded entangled photon source is when subjected to losses and dephasing due to naturally occurring interaction with its surrounding host material. 

In this respect, Axt \emph{et al.} investigated within a generating functions approach in Ref.~\onlinecite{Axt:PhysRevB:05} the dephasing of an exciton-biexciton-QD system, which is coupled to an arbitrary number of phonon modes and excited by a sequence of $\delta$-shaped classical light pulses, but did not consider the emission of single or entangled photons. With focus on photon entanglement Hohenester \emph{et al.} considered in Ref.~\onlinecite{Hohenester:PhysRevLett:07} the elastic phonon scattering at the device boundaries on a master equation level, assuming an asymmetry in the phonon coupling for different exciton states. 

In this paper we present microscopic calculations of a phonon-assisted biexciton cascade in an InAs QD embedded in a InGaAs wetting layer (WL).\cite{Bimberg::99} The coupling of the discrete QD states to the WL continuum via multi-phonon processes\cite{Inoshita:PhysRevB:92} leads to dephasing rates that significantly limit the entanglement output efficiency for high temperatures (above 100~K). 

The coupled dynamics of occupation densities and photon-assisted states in the two-photon emission is treated with an equation of motion (EOM) approach, well-established in solid state optics to approach many-particle problems.
This allows us to perform a quantum state tomography and thus calculate the concurrence of the polarization-entangled photons with varying, experimentally controlled external parameters like temperature, exchange splitting, or WL carrier density.

 The paper is organized as follows: After the considered system and calculated observables are introduced in more detail in Sec.~\ref{sec:qulight}, the different interactions of the QD carriers with phonons and cavity photons are discussed in Sec.~\ref{sec:model}. Then the coupled equations of motion are derived, their dynamics solved and the results presented in Sec.~\ref{subsec:eom} and~\ref{sec:results}.
\section{Characterization, generation, and measurement of entangled photons}\label{sec:qulight}
\subsection{Entangled photons and qubits}
The two quantum states ($\ket{0}$ and $\ket{1}$) that form a qubit $\qq$ can encode more information than a classical bit, as a qubit generally exists in a superposition 
$
\qq=c_1\ket{0}+c_2\ket{1}
$
with complex coefficients $c_{1}$ and $c_{2}$. In the model regarded here, the physical representation of the qubit states is the superposition of the horizontal $\hh$ and vertical $\vv$  polarization of photons.
The true advantage of the quantum approach is reached when two photonic qubits $Q=A,B$ interfere in a non-separable fashion and thus become entangled.\cite{Nielsen::00} 
The wave function of any entangled qubit $\ket{\psi}_Q$ can be written in the basis of the so-called Bell-states:\cite{Bell:Physics:64}
\begin{align*}
\ket{\Phi^\pm} 
&= 
\frac{1}{\sqrt{2}} 
 \left(
  \ket{HH} \pm \ket{VV}
 \right),\\
\ket{\Psi^\pm}
&= 
\frac{1}{\sqrt{2}} 
 \left(
  \ket{HV} \pm \ket{VH}
 \right),
\end{align*} 
where $\ket{HH}=\ket{H_AH_B}=\ket{H}_A\ket{H}_B$, similarly for $\ket{HV}, \ket{VH}, \ket{VV}$. Here, the entangled wave function cannot be expressed as a direct product $\ket{\psi}_Q 
\neq \ket{\psi_A}_Q \otimes \ket{\psi_B}_Q$ of the wave function of each single qubit. Before going into details of our analysis, we note, that the final photon wave function generated by a biexciton cascade QD emission can be expressed in terms of $\ket{\Phi^\pm}$ only (see discussion below or Ref.~\onlinecite{Benson:PhysRevLett:00}):
\begin{align*}
|\psi\rangle_Q 
&= 
 \frac{e^{\ii\theta}}{2} 
 \left( \ket{\Phi^+} - \ket{\Phi^-} \right)
+\frac{1}{2}
 \left( \ket{\Phi^+} + \ket{\Phi^-}  \right) \\
&= \frac{1}{\sqrt{2}} 
  \left( 
  \ket{HH} + e^{\ii\theta} \ \ket{VV}
  \right),
\end{align*}
where $\theta$ is the phase difference between $\ket{HH}$ and $\ket{VV}$. Both systems are in a complete statistical mixture between $\hh$ and $\vv$, but the coherence between $\ket{HH}$ and $\ket{VV}$ is kept. 
In the $HH, HV, VH, VV$ basis, the two-photon photon density matrix $\rho^\text{pt} = \ket{\psi}_{Q\;Q}\!\bra{\psi}$ of the pure $\ket{\psi}_Q$ state can be written as:
\begin{eqnarray}
\rho^\text{pt} &=& \ket{\psi}_{Q\;Q}\!\bra{\psi} =
 \begin{pmatrix}
   1 & 0 & 0 & e^{-\ii\theta} \\
   0 & 0 & 0 & 0 \\
   0 & 0 & 0 & 0 \\
   e^{\ii\theta} & 0 & 0 & 1 
 \end{pmatrix}. \label{eq:rho_qubit}
\end{eqnarray}
Following the Peres criterion,\cite{Peres:PhysRevLett:96} Eq.~\eqref{eq:rho_qubit} indicates entanglement since despite the remaining degree of freedom in $\theta$, the off-diagonal elements representing the coherence between the basis states can never be zero. If, however, these crucial matrix elements $\rho_{VH} = \bra{VV}\rho^{\text{pt}}\ket{HH}$ are zero, the system is said to have classical correlation only.

Since we consider entangled photons generated by optical transitions in a biexciton cascade, we introduce the QD's electronic band structure and selection rules in the next subsection.
	\subsection{Finestructure of quantum dots and generation of entangled photons}\label{subsec:bandstructure}
Depending on the intensity, photo-excitation creates in direct gap semiconductors electrons in the WL conduction band and holes in the WL valence band. The carriers subsequently relax into the QD and occupy the discrete energy shells following Pauli's principle.\cite{Stier:PhysRevB:99} Although relaxed, they still can interact with the WL via multi-phonon processes. To analyze entangled photon emission from a QD, its electronic eigenstates and selection rules for the light-matter interaction have to be known. 

\begin{figure*}[bt!]
\centering
\includegraphics[clip, width=\textwidth]{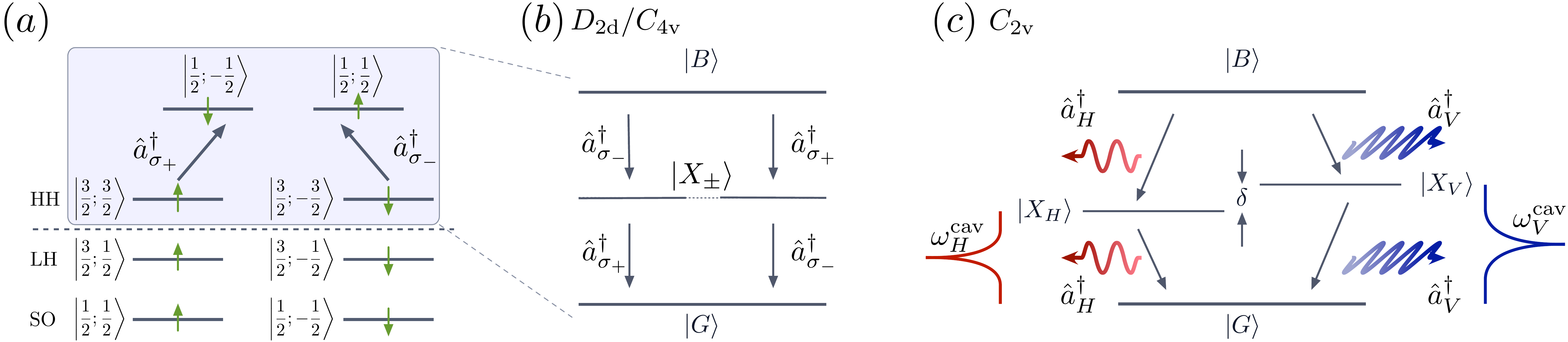}
\caption{(a) QD band structure with spin states $s = \ket{J;m_{j}}$. Relevant for the biexcitonic framework are only the electron and heavy hole (HH) states. (b) cascade of a symmetric QD with degenerate exciton states $|X_\pm\rangle$.  (c) cascade with a FSS $\delta \neq 0$ (without biexcitonic shift). There are two possible paths. Either two photons with vertical $V$ or horizontal $H$ polarization are emitted into a cavity mode $\omega^\text{cav}_{V/H}$ with FWHM $\kappa$. \\
For vanishing FSS $\delta$ the which-path information is lost and the photons are polarization-entangled. With a FSS $\delta \neq 0$ the photons are not fully polarization-entangled. } \label{fig:non_local_cascade_scheme}
\end{figure*}

Typically, when following a $\vec k \cdot \vec p$ approach, the bandstructure of single-InAs QDs around the $\Gamma$-point can be approximately described by the two anti-binding $s$-states of the electronsand the six binding $p$-states of the holes.\cite{Yu::05} Due to spin-orbit coupling and strain effects the split-off (SO) and light-hole (LH) states lie energetically well below the heavy-hole (HH) state. Therefore, the relevant single-particle basis is constructed by the heavy-hole with total angular momentum $J_{h}=3/2$ and spin projection in growth-direction $m_{j,h}=\pm3/2$ and the electron with $J_{e}=1/2$, $m_{j,e}=\pm1/2$, all shown in Fig.~\ref{fig:non_local_cascade_scheme}(a). The mutual Coulomb interaction will bind the carriers to electron-hole pairs and so lead to the formation of excitons. Depending on the configurations, given by the projections of the total angular momentum $M=m_{j,e}+m_{j,h}$ four exciton states arise, which can be characterized as optically inactive $M=\pm2$ (parallel electron and hole spin) and optically active $M=\pm1$ (anti-parallel spins).\cite{Efros:PhysRevB:96} In a $D_{2\text{d}}$-symmetric QD these bright exciton states, denoted by $|X_{\pm}\rangle$ are degenerated and couple to $\sigma_{\pm}$ circular polarized light [here, $\sigma_{+}$ ($\sigma_{-}$) labels right-hand (left-hand) circular polarized light] as depicted in Fig.~\ref{fig:non_local_cascade_scheme}(b).\cite{Weisbuch::91,Scholes:JChemPhys:04}    

The atom-like discrete energy levels of a QD can be pumped electrically or by photo-excitation.\cite{Yuan:Science:02} When the carriers occupy  their lowest shell, two excitons can form a bound singlet state, a biexciton ($\bb$). Under emission of the first, $\sigma_{+}$ or $\sigma_{-}$ polarized so-called biexciton-photon the system enters into an intermediate, optically allowed exciton state ($|X_{\pm}\rangle$). Subsequently, the QD relaxes into its ground state ($\GG$) by emitting a second, the exciton-photon.  As a consequence of total angular momentum conservation both emitted photons are of opposite circular polarization,\cite{Lochmann:ElectronLett:09} see Fig.~\ref{fig:non_local_cascade_scheme}(b). Since the exciton states $|X_{\pm}\rangle$ in symmetric $D_{2\text{d}}$ QDs are degenerate (i.e. no fine-structure splitting (FSS) $\delta$ is present), the photons' decay path can only be determined by their polarization, otherwise they are indistinguishable. Thus, the cascade will produce a maximally polarization-entangled photon pair wave function
$
\ket{\psi}_\text{pt}=(\ket{HH} + \ket{V V})/\sqrt{2}
$, 
which corresponds to the $\ket{\Phi^{+}}$ Bell-state.\cite{Edamatsu:JpnJApplPhys:07}

Although growth techniques of single QDs are very sophisticated,\cite{Seguin:ApplPhysLett:06} it is rarely possible to not have an asymmetry in the semiconductor crystal and so the non-classical correlation of the photons is often spoiled:  Under strain the dots symmetry reduces to $C_{2\text{v}}$ and the anisotropic electron-hole exchange interaction splits the exciton doublet into two states $\xh=1/\sqrt{2}(|X_{+}\rangle+|X_{-}\rangle)$ and $\xv=1/\sqrt{2}(|X_{+}\rangle-|X_{-}\rangle)$, energetically separated by the FSS $\delta$, shown in Fig.~\ref{fig:non_local_cascade_scheme}(c). These states couple to photon modes of orthogonal linear polarization along the direction of one crystallographic axis, labeled $H$ and $V$ respectively. This superimposes a which-path information onto the emitted photon frequencies and the degree of their entanglement is reduced. To efficiently collect the photons, the QD is placed inside a cavity supporting only two modes of different polarizations $V, H$ with frequency $\omega^\text{cav}_{V}\neq\omega^\text{cav}_{H}$.\cite{Hennessy:ApplPhysLett:06} These modes are assumed to be in resonance with the corresponding exciton-ground state transitions. Although energetically off-resonant, the biexciton-photon is emitted into the same mode, see Fig.~\ref{fig:non_local_cascade_scheme}(c). In principal, the exciton states can be tuned into near resonance again by applying an in-plane external electric \cite{Reimer:PhysRevB:08, Korkusinski:PhysRevB:09} or magnetic field,\cite{Stevenson:PhysRevB:06} and indeed for a small FSS $\delta$ generation of polarization-entangled photon pairs was demonstrated on a system operating at 10~K.\cite{Stevenson:Nature:06}

Although the ideal case of zero splitting can be recovered, phonons as a decoherence mechanisms, in particular at elevated temperatures, will have an impact on the performance of a QD as a source of polarization-entangled photons. To provide a meaningful quantitative measure of the entanglement the next section will introduce the concurrence.
	\subsection{Measure of entanglement -- relevant quantities}
As shown in Eq.~\eqref{eq:rho_qubit}, a measure for the degree of entanglement is determined by the off-diagonal element in the polarization sub-space of the two-photon density matrix $\rho^\text{pt}$, explicitly:
\bqn
\rho_{VH} := \langle V V | \rho^\text{pt} | H H \rangle .
\eqn
Quantum state tomography \cite{James:PhysRevA:01} provides a measurement scheme which gives access to the different elements of $\rho^\text{pt}$. They are experimentally reconstructed by measuring of the two-photon cross correlation function $g^2_{ij}(t,\tau)$ \cite{Mandel::95} over a mean photon arrival time $t$. The function $g^2_{ij}(t,\tau)$ corresponds to the polarization correlation between a biexciton-photon emitted at time $t$ and the subsequent exciton-photon at time $t+\tau$:\cite{Troiani:PhysRevB:06}
\bqn
g^2_{ij}(t,\tau) \propto 
\ew{a^\dg_{i}(t) a^\dg_i(t+\tau) a^\ndg_j(t+\tau) a^\ndg_j(t)},\label{eq:g2}
\eqn 
where $i,j \in \lbrace H,V \rbrace$. The correlation function is written in terms of photon creation $a^\dg_{i}$ and annihilation $a^\ndg_{j}$ operators of the different photon modes $i,j$, cp. Fig.~\ref{fig:non_local_cascade_scheme}(c).
We consider an experimental setup, where the time delay between the two photons $\tau$ is zero. This can be realized by appropriately adjusting the distance to the detector.\cite{Stevenson:Nature:06, Benyoucef:NewJPhys:04} The temporal dynamics of the corresponding density matrix element $\langle ii | \rho^\text{pt} | jj \rangle$ is obtained when the second-order correlation function $g^2_{ij}(t,\tau=0)$ is time averaged over the arrival times $t$:\cite{Troiani:PhysRevB:06}
\bqn
\langle ii | \rho^\text{pt} | jj \rangle (t)
= 
\frac{1}{t}\int_0^t  g^2_{ij}(t',0) \ \dint t' .
\eqn
Thus, the source of entanglement can be rewritten as $\rho_{HV}\propto\ew{ \hat a^\dg_{V} \hat a^\dg_{V} \hat a^\ndg_{H} \hat a^\ndg_{H}}$. A standard expression for the degree of entanglement is the concurrence $C$:\cite{Wootters:PhysRevLett:98,Coffman:PhysRevA:00}
\bqn
C = 2 
\vert \rho_{HV} \vert, \label{eq:concurrence}
\eqn
directly related to other accepted measures like the entanglement of formation\cite{Bennett:PhysRevA:96} $E_{F}$ or the tangle\cite{White:PhysRevA:01} $T$ (for example, $C=\sqrt{T}$). Here, $C=1$ corresponds to maximum entanglement and $C=0$ to zero entanglement.
To calculate the necessary dynamics of the expectation values, e.g. $\ew{ \hat a^\dg_{V} \hat a^\dg_{V} \hat a^\ndg_{H} \hat a^\ndg_{H}}$, a set of equations is derived in the next section.
\section{Modeling an embedded quantum dot as a semiconductor source of entanglement}\label{sec:model}
	\subsection{Hamiltonian and QD model}
The coupled dynamics of observables, such as Eq.~\eqref{eq:g2} can be generally derived from the system's Hamilton operator $\hat H$ via Heisenberg's equation of motion:
\bee
-\ii\hbar\partial_t\ew{\hat O}=\ew{[\hat H,\hat O]_{-}}.\label{eq:heisenberg}
\ene
In this section, we discuss the used Hamiltonian to implement our model system. Applying Eq.~\eqref{eq:heisenberg} to the photon correlation function $g^{2}_{ij}$ introduces a coupling to the electronic degrees of freedom via the electron-photon interaction. As discussed in Sec.~\ref{subsec:bandstructure}, due to strong confinement, we assume that the QD single-particle states are energetically well separated and only the highest valance ($v$) HH state and lowest conduction ($c$) band $s$-shell form the biexcitonic framework, see again Fig.~\ref{fig:non_local_cascade_scheme}. In second quantization Fermionic operators describe the typical electron-hole representation, where carriers are the heavy holes in the $v$ band (operator $\hat h_{s}$) and electrons in the $c$ band (denoted by operator $\hat e_{s}$). Here, the carrier spin state $s=\ket{J,m_j}$ is given for the holes (electrons) by $\ua=\ket{3/2;3/2}$ ($\ua=\ket{1/2;1/2}$) and $\da=\ket{3/2;-3/2}$ ($\da=\ket{1/2;-1/2}$).

\begin{figure}[tb]
\centering
\includegraphics[clip, width=\columnwidth]{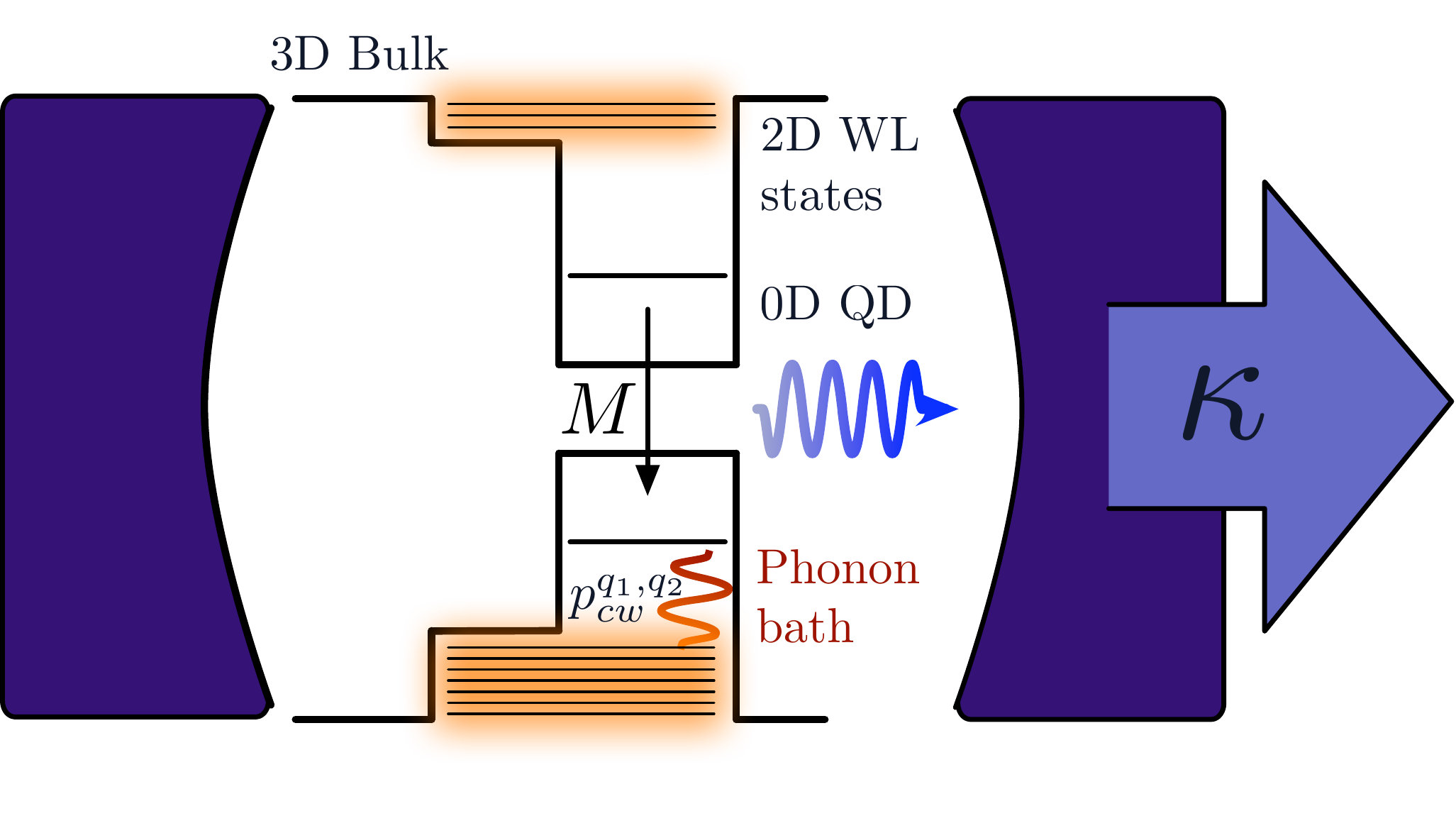}
\caption{General system scheme. Two electron-hole pairs in a QD are coupled to the continuum of WL states via LO-phonon interaction $p^{q_1,q_2}_{cw}$. The phonons are in a thermal bath. \\
The QD has two levels with a conduction and a valence band. Here, the carriers interact with photons via the electron-light coupling elements $M$. The QD is assumed to be placed at a node position of the electromagnetic field in a cavity. Since its loss $\kappa=10~\mu$eV is greater then the coupling strength to the field $ M=1~\mu$eV, the system is in a weak coupling regime.} \label{fig:local_system_scheme}
\end{figure}

The complete Hamiltonian of a QD coupled to the WL continuum inside a nanocavity consists of various parts and is given as:
\begin{align}
\hat H &= \hat H^\text{c}_{\text{QD},0} +\hat  H^\text{c-c}_\text{QD} + \hat H^{\text{pt}}_\text{0} + \hat H^\text{c-pt}_\text{QD}\nt\\
 &+ \hat H^\text{c}_{\text{WL},0}  + \hat H^{\text{pn}}_\text{0} + \hat H^\text{c-pn}_\text{QD,WL} + \hat H^\text{c-pn}_\text{WL,WL}.\label{eq:fullHam}
\end{align}
First, the kinetic energy of the confined QD carriers $\hat H^\text{c}_{\text{QD},0}$ and their mutual carrier-carrier interaction $\hat  H^\text{c-c}_\text{QD}$ are introduced. The electron-hole pair in the QD does interact with the cavity photons of the quantized light field $\hat H^\text{c-pt}_\text{QD}$ and so the free energy of the cavity photons $\hat H^{\text{pt}}_\text{0}$ is included as well. The free energy of semiconductor bulk phonons $\hat H^{\text{pn}}_\text{0}$ and WL carriers $\hat H^\text{c}_{\text{WL},0}$ appears, too. The interaction of the WL with the QD states via LO-phonons is considered in $\hat H^\text{c-pn}_\text{QD,WL}$ and the electron-phonon coupling within the WL in $\hat H^\text{c-pn}_\text{WL,WL}$. The contributions to the $\hat H_{0}$ part read:
\begin{align*}
\hat H^\text{c}_{\text{QD},0} =& \sum_{s} \varepsilon^\text{QD}_{v} \erzqd{h}{s}\verqd{h}{s}+\varepsilon^\text{QD}_{c} \erzqd{e}{s}\verqd{e}{s}, \\
 \hat H^{\text{pn}}_{0}=&\sum_{\vec q}\hbar\omega_\text{LO}\hat b^\dg_{\vec q}\hat b^\ndg_{\vec q},\\
\hat H^\text{c}_\text{WL, 0} =& \sum_{\vec k s} \varepsilon^\text{WL}_{v \vec k} \erzqd{w}{\vec k s}\verqd{w}{\vec k s},\\
 \hat H^{\text{pt}}_\text{0}=&\sum_{i}\hbar\omega_{i}\hat a^\dg_{i}\hat a^\ndg_{i}.
\end{align*}
The Bosonic longitudinal optical (LO) phonon creation (annihilation) operators at wave vector $\vec q$ are $\hat b^\dagger_{\vec q}$ ($\hat b^\ndg_{\vec q}$). Their dispersion is treated within the Einstein approximation and $\hbar\omega_\text{LO}= 36$~meV. Similar to the QD operators $\hat w_{\vec ks}^\dagger$ ($\hat w_{\vec ks}^\ndg$) are creators (annihilators) of a hole carrier in the WL continuum of the valence band $v$ with spin state $s$ and wave vector $\vec k$. For the WL carriers, we take into account only the hole contributions of the $v$ band, motivated in the next section. The impact of spin-orbit coupling on the carrier's energy can be neglected in QDs~\cite{Vachon:PhysRevB:09} and so, $\varepsilon_i$ is assumed to be independent of the carrier's spin state. The general considered setup is displayed in Fig.~\ref{fig:local_system_scheme}.
The next subsections discuss the remaining parts of the total Hamiltonian  separately and in more detail. The electron-phonon interaction Hamiltonian leads to temperature dependent dephasing rates (see subsection~\ref{sec:wetting-layer-hamiltonian}). The pure electronic part of non-interacting and Coulomb-interacting QD electrons $\hat H^\text{c}_{\text{QD},0} +\hat  H^\text{c-c}_\text{QD}$ is diagonalized and transformed into an excitonic basis. The new arising eigenvalues and eigenvectors will incorporate the complete Coulomb contributions (see subsection~\ref{sec:coulomb-hamiltonian}), which are energy shifts (e.g. ground state and biexciton shift) and the exchange splitting due to different spin combinations in the exciton states. In the excitonic basis, the electron-photon interaction is transformed and new orthogonal field modes are derived (see subsection~\ref{sec:electron-light-hamiltonian}). 
		\subsubsection{Multi-phonon coupling of WL and QD}\label{sec:wetting-layer-hamiltonian}
Embedded in a host material, quantum confined electrons in Stranski-Krastanov grown InAs/GaAs QDs interact via LO-phonons with a continuum of two dimensional electronic WL states only a few ten meV away. This leads to temperature dependent dephasing times for the QD states \cite{Magnusdottir:JApplPhys:02,Muljarov:PhysRevLett:07,Muljarov:PhysStatusSolidiB:08,Seebeck:PhysRevB:05} and can be a first approach to the problem of temperature dependent generation of entangled photon pairs. On the time scale of several nanoseconds regarded here, longitudinal acoustical phonons\cite{Machnikowski:PhysRevB:08,Krummheuer:PhysRevB:05} only have a minor impact and are not considered. 

\begin{figure}[bt]
\centering
\includegraphics[clip, width=\columnwidth]{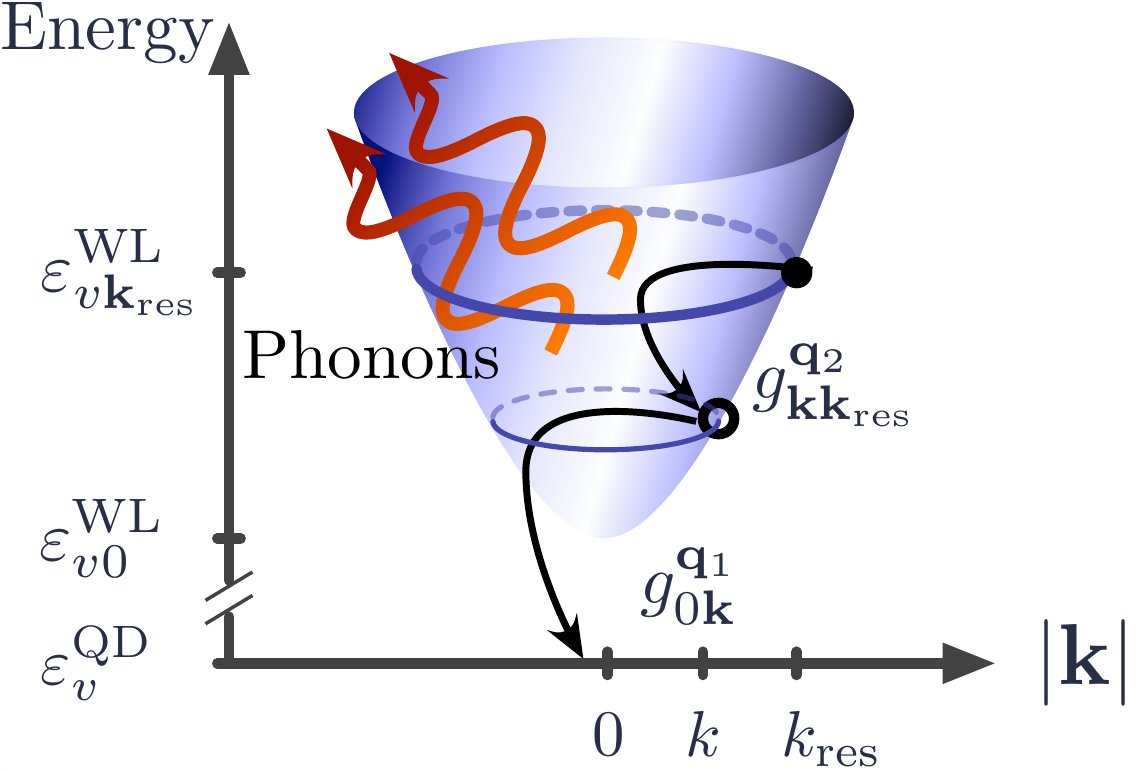}
\caption{Wetting layer system scheme. Seen in blue is the 2D dispersion of the WL holes $\hbar^2\vec k^2/({2m_{h}^*})$. Marked on the energy scale are (i) the WL band edge $\varepsilon^{\text{WL}}_{v 0}$ which is energetically separated from the QD $v$ shell $\varepsilon_v^\text{QD}$ by more then a single-phonon energy:  $\Delta\varepsilon^{\text{QD}}_{v}=\varepsilon^{\text{WL}}_{v 0}-\varepsilon_v^\text{QD}>\hbar\omega_\text{LO}$, (ii) the WL states at $\vec k_\text{res}$ resonant with a two-phonon transition $\varepsilon^{\text{WL}}_{v \vec k_\text{res}}=\varepsilon^{\text{QD}}_{v}+2\hbar\omega_\text{LO}$ to the QD.\\
The transition from the resonant states to the QD pass through intermediate states at $\varepsilon^{\text{WL}}_{v \vec k}$ with probability amplitude $g^{\vec q}_{\vec k_2 \vec k_1}$. } \label{fig:mrd_scheme}
\end{figure}

\begin{figure}[bt!]
\centering
\includegraphics[clip, width=\columnwidth]{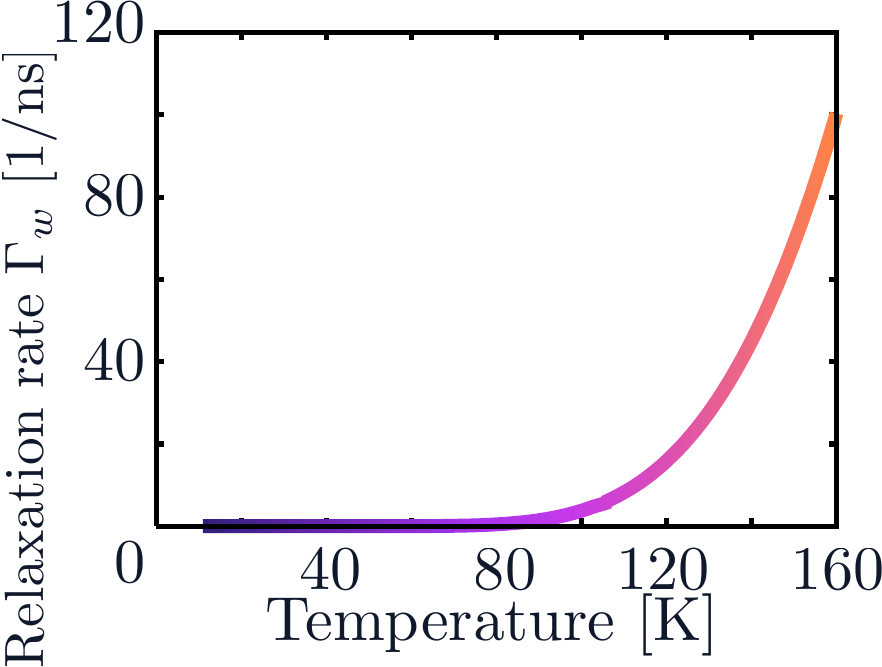}
\caption{Plot of the phonon-induced relaxation rate $\Gamma_{w}$ as a function of temperature $T$. At about 80~K $\Gamma_{w}$ has approached 1~ns$^{-1}$ and starts to contribute strongly to the decay of the QD states.} \label{fig:mrd_rated}
\end{figure}

Depending  on the dot size,\cite{Stier:PhysRevB:99} the QD valence band is typically more than one, but less than two LO-phonon energies $\hbar\omega_\text{LO}$ energetically separated from the WL band edge, see Fig.~\ref{fig:mrd_scheme}. To effectively connect the QD states with the WL, up to two-phonon processes have to be taken into account. Within the two-phonon limit the influence of the WL conduction band on the QD electrons can be neglected, because here, more than two LO-phonons are necessary to bridge the energy gap to the WL and therefore the dephasing is determined by the hole-WL interaction. Moreover, the Coulomb interaction of the WL carriers is not included as the carrier densities considered here are low.\cite{Lorke:PhysRevB:06} Under these assumptions, microscopic dephasing rates are derived by using an effective Hamiltonian approach, which originates from a multi-photon theory.\cite{Faisal::87}
From the Hamiltonian in Eq.~\eqref{eq:fullHam} the following parts contribute to the LO-phonon induced dephasing:
\begin{align}
\hat H_\text{QD,WL}=&\hat H^\text{c}_{\text{QD},0}+ \hat H^\text{c}_{\text{WL},0}\nt \\
  +& \hat H^{\text{pn}}_\text{0} + \hat H^\text{c-pn}_\text{QD,WL} + \hat H^\text{c-pn}_\text{WL,WL}.
\label{eq.el-ph.hamilton}
\end{align}
The phonon mediated interaction between QD holes and WL states and the carrier-phonon interaction within the WL are given by
\begin{align*}
\hat H^\text{c-pn}_\text{QD,WL}&=\sum_s\sum_{\vec k \vec q}{} g_{0\vec k}^{\vec q} \hat h_{s}^\dg \hat w_{\vec ks}^\ndg(\hat b_{\vec q}^\ndg+\hat b^\dg_{-\vec q})+ \text{h.a.},
\\
 \hat H^\text{c-pn}_\text{WL,WL}&=\sum_s\sum_{\vec k \vec k' \vec q}{} g_{\vec k\vec k'}^{\vec q}\hat w_{\vec ks}^\dg \hat w_{\vec k's}^\ndg(\hat b_{\vec q}^\ndg+\hat b^\dg_{-\vec q})+ \text{h.a.}.
\end{align*}
The Fr\"ohlich coupling elements are $g_{\vec k\vec k'}^{\vec q}$ which can be found in Ref.~\onlinecite{Mahan::90}. Within a projection operator based  theory \cite{Faisal::87} Eq.~\eqref{eq.el-ph.hamilton} is mapped onto the resonant WL states only and becomes: 
\begin{equation}
\hat H_\text{QD,WL}=\hat H^\text{c}_{\text{QD},0}+ \hat H^\text{c}_{\text{WL},0} + \hat H_{\text{eff}},\label{eq:heff}
\end{equation}
with the effective LO-phonon-assisted WL influence on the QD holes in $\hat H_{\text{eff}}$. All other off-resonant contributions are implicitly included in the coupling elements of the effective Hamiltonian. Taking only two-phonon processes into account, $\hat H_{\text{eff}}$ reads:\cite{Dachner:PhysStatSolBsubmitted:09}
\begin{align*}
\hat H_{\text{eff}}
=& \sum_s\sum_{\vec q_1 \vec q_2  \vec k_{\text{res}} }p_{cw,s}^{\vec q_1\vec q_2} \hat h_{s}^\dagger \hat w_{\vec k_{\text{res}}s}^\ndg \hat b_{\vec q_1}^\ndg \hat b_{\vec q_2}^\ndg \\
+& \sum_s\sum_{\vec q_1 \vec q_2  \vec k_{\text{res}}}p_{wc,s}^{\vec q_1\vec q_2}\hat w_{\vec k_{\text{res}}s}^\dagger \hat h_{s}^\ndg \hat b^\dagger_{\vec q_1} \hat b^\dagger_{\vec q_2},
\end{align*}
with the effective coupling elements
\begin{equation}
 p_{cw,s}^{\vec q_1\vec q_2}=\sum_{\substack{\vec k\\\vec k\neq \vec k_\text{res}}}\frac{g^{\vec q_1}_{0\vec k}g^{\vec q_2}_{\vec k\vec k_\text{res}}
(1-\langle \hat w_{\vec ks}^\dagger \hat w_{\vec ks}^\ndg\rangle)}{\varepsilon^{\text{WL}}_{v \vec k}-\varepsilon^{\text{QD}}_{v}-\hbar\omega_\text{LO}}.\label{eq:eff_couple}
\end{equation}
They contain Pauli blocking terms and therefore depend on temperature and WL carrier occupation $(1-\langle \hat w_{\vec ks}^\dagger \hat w_{\vec ks}^\ndg\rangle)$.\cite{Wolters:PhysRevBsubmitted:09} The WL holes in $\hat H_{\text{eff}}$ have an energy exactly two phonon energies away from the QD state energy. 
A transition from these resonant WL holes at $\varepsilon^{\text{WL}}_{v \vec k_\text{res}}$ to the QD shell takes place under simultaneous emission of two phonons. The whole process is energy conserving: $\varepsilon^{\text{WL}}_{v \vec k_\text{res}}-\varepsilon^{\text{QD}}_{v}=2\hbar\omega_\text{LO}$. Within time-energy uncertainty carriers relax by a higher-order Markov process. Here, in the transition to the intermediate state at $\varepsilon^{\text{WL}}_{v \vec k}$ energy conservation is violated, since the hole state at $\vec k$ is less than $\hbar \omega_\text{LO}$ from $\varepsilon^{\text{WL}}_{v \vec k_\text{res}}$ and more than $\hbar \omega_\text{LO}$ from the QD state energetically separated, see Fig.~\ref{fig:mrd_scheme}.
The probability amplitude for the intermediate transitions are $g^{\vec q_2}_{\vec k\vec k_\text{res}}$ and $g^{\vec q_1}_{0\vec k}$. 
Equation~\eqref{eq:eff_couple} shows that all possible transitions between QD and WL are mediated by all off-resonant WL states $\vec k$. The strength of the coupling element is determined by to what extent the energy conservation condition in the denominator is met in every phonon-assisted electronic transition.
We can use the effective Hamiltonian Eq.~\eqref{eq:heff} to derive relaxation and dephasing rates using Heisenberg equations of motion, where the hierarchy problem is treated within a born factorization.\cite{Waldmueller:IEEEJQuantumElectron:06} The calculations lead to the following equations for the QD states:
\begin{eqnarray}
 \frac{d}{dt}\langle \hat e_{s}^\dagger \hat h_{s}^\dg\rangle
&=&-(\ii\hbar\Omega_0+\Gamma_{w,s})\langle \hat e_{s}^\dagger \hat h_{s}^\dg\rangle,
\label{eq.qdwl.ratengl.pol} %
\\
\frac{d}{dt}\langle \hat h_{s}^\dagger \hat h_{s}^\ndg\rangle
&=&-2\Gamma_{w,s}\langle \hat h_{s}^\dagger \hat h_{s}^\ndg\rangle,
\label{eq.qdwl.ratengl.dens}
\end{eqnarray}
with the QD gap energy $\hbar\Omega_0=\varepsilon^{\text{QD}}_c-\varepsilon^{\text{QD}}_v$ and the WL-induced damping rate:\cite{Dachner:PhysStatSolBsubmitted:09}
\begin{equation}
\Gamma_{w,s}=\left[(1-f_{h,s})(n+1)^2+f_{h,s} n^2   \right]\gamma_{s}.\label{eq:dampingRates}
\end{equation}
The damping $\gamma_s$ is given by:
\begin{align}
\gamma_s =& \iint \text{d}^3q_1\text{d}^3q_2\quad p_{cw,s}^{\vec q_1\vec q_2}(p_{wc,s}^{\vec q_1\vec q_2}+p_{wc,s}^{\vec q_2\vec q_1}) .
\end{align}
In Eq.~\eqref{eq:dampingRates} $f_{h,s}=\sum_{\vec k_\text{res}}\langle \hat w_{\vec k_\text{res} s}^\dagger \hat w^\ndg_{\vec k_\text{res}s}\rangle$ is used for the WL hole density at the resonant energy, which in the carrier low-density limit is assumed to be zero. Note, that this implies $\Gamma_{w}=\Gamma_{w,\ua}=\Gamma_{w,\da}$. Since the system has relaxed into a quasi-equilibrium, the phonon bath is described by the Bose-Einstein distribution $n=\langle \hat b^\dagger \hat b\rangle$. 
Figure~\ref{fig:mrd_rated} displays the temperature dependence of $\Gamma_w$.
In the next section an excitonic basis will be introduced. The damping rates in Eqs.~\eqref{eq.qdwl.ratengl.pol} and~\eqref{eq.qdwl.ratengl.dens} will be used in this basis to account for the relaxation of the QD electrons.\cite{Dachner::09} 
		\subsubsection{Carrier-carrier interaction and exciton representation}\label{sec:coulomb-hamiltonian}
The Hamiltonian in Eq.~\eqref{eq:fullHam} accounting for the QD carriers and their interaction via the Coulomb potential $\hat H_\text{Coul}=\hat H_{\text{QD},0}^{\text{c}} +\hat  H^{\text{c-c}}_\text{QD}$ is conveniently rewritten as:\cite{Axt:PhysRevB:05}
\begin{align}
\hat H_\text{Coul}
&=\phantom{\frac{1}{2}}\sum_s \ 
      \frac{\hbar\Omega_0}{2} 
      \left( \hat e^\dg_{s}\hat e^\ndg_{s}
            -\hat h^\dg_{s}\hat h^\ndg_{s}  
      \right)   \label{eq:H_Coulomb} \\ 
\nonumber
 &+\frac{1}{2}\sum_{ss'} \Bigl( \phantom{+}
V^{ee} \ \hat e^\dg_{s} \hat e^\ndg_{s} 
      \hat e^\dg_{s'} \hat e^\ndg_{s'} 
+ 
V^{hh} \ \hat h^\dg_{s} \hat h^\ndg_{s} 
      \hat h^\dg_{s'} \hat h^\ndg_{s'}\nt \\ 
&\phantom{+\frac{1}{2}\sum_{ss'}} \ + 2
V^\text{ex}_{ss'} \ \hat e^\dg_{s} \hat 
      \hat h^\dg_{s} h^\ndg_{s'} \hat e^\ndg_{s'}
-2 V^{he} \ \hat e^\dg_{s} \hat e^\ndg_{s} 
        \hat h^\dg_{s'} \hat h^\ndg_{s'}\Bigr)  . \nt
\end{align}
The first term contains the non-interacting electrons with gap energy $\hbar\Omega_0$. The second term accounts for the repulsion of carriers within the same band, whereas the last term gives attractive direct Coulomb interaction and repulsive exchange interaction between carriers in different bands. The corresponding Coulomb elements $V^{ee}, V^{hh}, V^{he}, V^\text{ex}_{ss'}$ mediate the interaction. 
Responsible for the fine structure splitting $\delta$, compare Fig.~\ref{fig:non_local_cascade_scheme}, is the exchange splitting $V^\text{ex}_{\ua\da}$, which describes the repulsion and attraction forces induced by different spin-conformations of electrons and holes. As shown in App.~\ref{app:ccHam} the FSS can be expressed by
\bqn
\delta=2 \left|V^\text{ex}_{\ua\da}\right|.
\eqn 
To simplify the notation we will refer to $V^\text{ex}_{\ua\da}$ as $\vex$.

In principle, all matrix elements of the Coulomb interaction can be microscopically calculated, when the single-particle wave functions are known.\cite{Takagahara:PhysRevB:00,Takagahara:PhysRevB:93,Richter:PhysStatusSolidiB:06,Carmele:PhysRevB:09} However, their values only have a quantitative impact on the results. Therefore, within a reasonable range, they are used as model parameters, measured in experiments.\cite{Langbein:PhysRevB:04}
\begin{small}
\begin{table}
 \caption{\label{tab:param}Numerical parameters.}
\begin{ruledtabular}
  \begin{tabular}{lcc}
 Parameter&Symbol&Value
\\
\hline
electron effective mass&$m_e$& 0.043$m_0$ \cite{Rossler::2002}
\\
hole effective mass&$m_{h}$&0.450$m_0$ \cite{Rossler::2002}
\\
LO-phonon energy&$\hbar\omega_{\rm LO}$& 36.4~meV \cite{Rossler::2002}
\\
QD band gap&$\hbar \omega_0$&1.5~eV
\\
hole binding energy&$\Delta\epsilon_{v}$&1.5~$\hbar\omega_\text{LO}$
\\
Coulomb parameters&$V^{vc}=V^{vv}=V^{cc}$&25~$\mu$eV
\\
&$V^\text{ex}_{\ua\ua}=V^\text{ex}_{\da\da}$&$0~\mu$eV
\\
photon lifetime in a cavity&$\kappa$&10~$\mu$eV
\\
electron-photon coupling&$M$& 1~$\mu$eV 
\\
photon lifetime&$\Gamma_\text{rad}$&50~ps$^{-1}$ 
\\
\end{tabular}
\end{ruledtabular}
\end{table}
\end{small}
Using the space spanned by the new exciton operators, derived in App.~\ref{app:ccHam}, the Hamiltonian is rewritten as:\cite{Richter:PhysStatusSolidiB:06}
\begin{align*}
\hat H_\text{Coul} & = 
\hbar\omega_G \ \hat G^{\dg} \hat G^{\ndg}
+
\hbar\omega_H \ \hat X^{\dg}_H \hat X^{\ndg}_H\\
&+
\hbar\omega_V \ \hat X^{\dg}_V \hat X^{\ndg}_V
+
\hbar\omega_B \ \hat B^{\dg} \hat B^{\ndg},
\end{align*}
with $\hat G$ the ground state, $\hat X_{H/V}$ the exciton and $\hat B$ the biexciton annihilation operator, corresponding to the excitonic level structure as depicted in Fig.~\ref{fig:non_local_cascade_scheme}(c). Note, that within this diagonal representation the derivation of equations of motion via Eq.~\eqref{eq:heisenberg} is trivial for the Coulomb interaction, as operators of different states commute.
		\subsubsection{QD electron-photon interaction}\label{sec:electron-light-hamiltonian}
Commonly, the Hamiltonian of the electron-photon interaction is taken in rotating-wave approximation:\cite{Mandel::95}
\bqn
\hat H_{\text{QD}}^\text{c-pt} = \hbar M \sum_{i}
\left( \hat h^\ndg_{\ua} \hat e^\ndg_{\downarrow} \hat a^\dg_{i \sigma_{+}} 
+  
\hat h^\ndg_{\da} \hat e^\ndg_{\uparrow} \hat a^\dg_{i \sigma_{-}} \right)+ \text{h.a.},\label{eq:Helpt}
\eqn
where the corresponding spin states couple to circular $\sigma_{\pm}$ polarized light (see Sec.~\ref{subsec:bandstructure} for more detail), $i$ is the mode of the emitted photon and $M$ denotes the electron-photon coupling matrix elements. 
The Hamiltonian of the light-electron interaction is expressed with the exciton operators, derived in the App.~\ref{app:eptHam}:
\begin{eqnarray}
\hat H_{\text{QD}}^\text{c-pt} = &\phantom+& \hbar M \sum_{i}
\left(
\hat G^\dg \hat X_H^{\ndg} \ \hat a^\dg_{iH} 
+
\hat G^\dg \hat X_V^{\ndg}  \ \hat a^\dg_{iV} 
\right) \\
&+& \hbar M \sum_{i}
\left(
\hat X_H^{\dg} \hat B \ \hat a^\dg_{iH}
-
\hat X_V^{\dg} \hat B \ \hat a^\dg_{iV}
\right) 
 + \text{h.a.} .\nt
\end{eqnarray}
We assume that our QD is placed at a node position inside a nano-cavity, which provides two different modes for the different polarizations $H$ and $V$ corresponding to different frequencies $\omega_{H}\neq\omega_{V}$.\cite{Hennessy:ApplPhysLett:06} Since only two modes exist within the cavity we investigate a cavity-enhanced biexciton cascade. However, we remain in the weak coupling regime since the cavity loss $\kappa=10~\mu$eV is greater then the coupling strength to the field $ M=1~\mu$eV. Regarding only these two modes, the electron-light interaction Hamiltonian can be written in a compact form
\begin{align}
\hat H_{\text{QD}}^\text{c-pt} = \hbar M
\Bigl(
 \phantom{+}& \hat G^\dg \hat X_H^{\ndg} \ \hat a^\dg_{H} 
+
\hat G^\dg \hat X_V^{\ndg}  \ \hat a^\dg_{V} 
 \nt \\ 
+&  
\hat X_H^{\dg} \hat B \ \hat a^\dg_{H}
-
\hat X_V^{\dg} \hat B \ \hat a^\dg_{V}
\Bigr) 
 + \text{h.a.} .
\end{align}
At this status, the total Hamiltonian is written with the new exciton and photon operators ($H,V$). To discuss the polarization entanglement between two emitted photons we proceed and determine equations of motion, which govern the relevant observables' dynamics.
	\subsection{Equations of motion}\label{subsec:eom}
\begin{figure}[bt!]
\centering
\includegraphics[clip, width=\columnwidth]{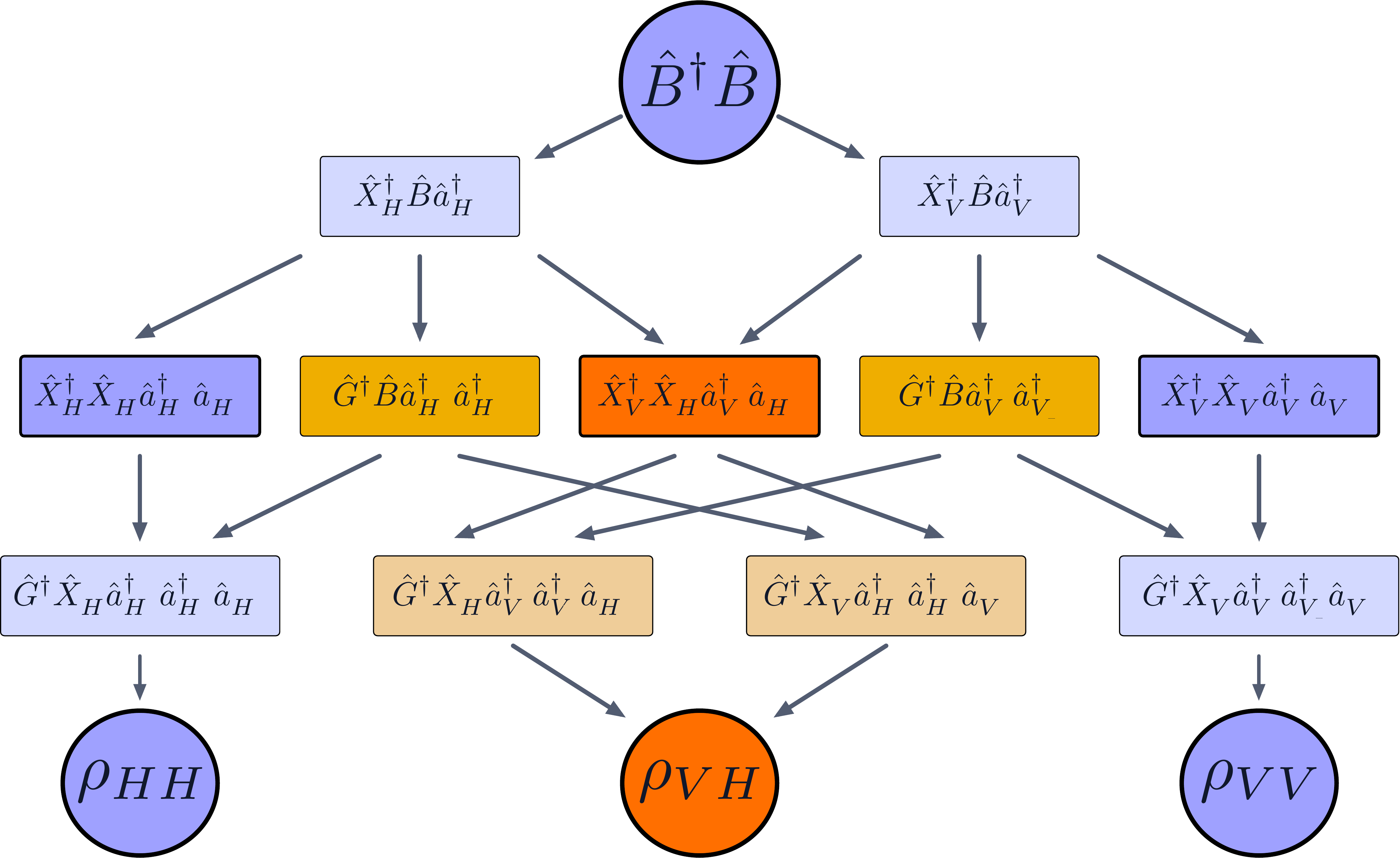}
\caption{Equations of motion, truncated to the pure cascade scheme. The dark blue quantities represent densities which do not contribute to entanglement, whereas the dark orange and red quantities directly generate a crossing of the different paths in the light orange boxes and are crucial to entanglement. } \label{fig:eom_scheme}
\end{figure}
To derive the coupled dynamics of photon-polarization coherences and electronic transitions with Heisenberg's equation of motion Eq.~\eqref{eq:heisenberg},\cite{Ahn:PhysRevB:05,Richter:PhysRevLett:09} the total excitonic Hamiltonian 
\begin{align*}
\hat H &= \sum\nolimits_{i=H,V}\hbar\omega^{\text{cav}}_{i}\erzqd{a}{i}\verqd{a}{i}
+ \hbar\omega_G \ \hat G^{\dg} \hat G^{\ndg} + \hbar\omega_H \ \hat X^{\dg}_H \hat X^{\ndg}_H \\
 &+  \hbar\omega_V\ \hat X^{\dg}_V \hat X^{\ndg}_V + \hbar\omega_B \ \hat B^{\dg} \hat B^{\ndg} 
	+\hbar M
\left(
\hat G^\dg \hat X_H^{\ndg} \hat a^\dg_{H} 
\right.\\
&+\left.
\hat G^\dg \hat X_V^{\ndg} \hat a^\dg_{V}+\hat X_H^{\dg} \hat B \hat a^\dg_{H}
-
\hat X_V^{\dg} \hat B \hat a^\dg_{V}
+ \text{h.a.}
\right) 
\end{align*}
in conjunction with the temperature dependent relaxation rates $\Gamma_w$ given in Eq.~\eqref{eq:dampingRates} is used. The latter lead to an additional decay of the QD populations and dephasing contributions to the QD transitions. Those contributions are derived via the effective Hamiltonian approach in Sec.~\ref{sec:wetting-layer-hamiltonian} and are consistently included by higher-order Markov approximations of the phonon-mediated interaction between carriers in the WL and QD.\cite{Dachner::09}

An overview of the, at a first glance complicated coupled dynamics of the considered correlation functions (derived by Eq.~\ref{eq:heisenberg}), is given in Fig.~\ref{fig:eom_scheme}. Going through the scheme, step by step we unravel the consequential interplay of the different quantities. An initially given biexciton density can decay via two possible paths (left $H$, right $V$) and generate a photon pair $\rho_{ij}$($i,j \in \lbrace H,V \rbrace$). In a first step, a photon-assisted coherence builds up $\hat X_i^{\dg} \hat B \ \hat a^\dg_{i}$ (light blue box), which then contributes to (i) a cross-polarization coherence $\hat X_V^{\dg} \hat X_H^{\ndg} \ \hat a^\dg_{V} \hat a^\ndg_{H}$ (red box) particularly important to achieve entanglement in $\rho_{VH}$. (ii) a two-photon coherence $\hat G^{\dg} \hat B^{\ndg} \ \hat a^\dg_{i} \hat a^\dg_{i}$ (light orange box), which also leads to an interference of the two path and thus contributes to $\rho_{VH}$. (iii) a combined exciton-photon density $\hat X_i^{\dg} \hat X_i^{\ndg} \ \hat a^\dg_{i} \hat a^\ndg_{i}$ (dark blue box), which does not influence the degree of entanglement. This gives meaningful insights to the underlying physics and origin of polarization entanglement. Note, that we are in a weak coupling regime and only spontaneous emission in the cascade is taken into account. \newline The concurrence $C$ as a measure for the degree of entanglement is determined by the photon density matrix, cf. Eq.~\eqref{eq:concurrence}, and defined via the off-diagonal element $\rho_{VH}$:
\begin{align}
\partial_t 
& \ew{\hat a^\dg_{V} \hat a^\dag_{V} \hat a^\ndg_{H} \hat a^\ndg_{H} }
= \label{eq:eomRhoVH}\\
 \phantom{+}&2\ii\ (\omega_V^\text{cav} - \omega_H^\text{cav} + 2\ii\kappa)
  \ew{\hat a^\dag_{V} \hat a^\dag_{V}  \hat a^\ndg_{H} \hat a^\ndg_{H}}
\nt \\
\nonumber
+&2\ii M \left(
\ew{ \hat G^\dg \hat X^{\phdg}_H \hat a^\dg_{V} \hat a^\dag_{V} \hat a^\ndg_{H}}
-
 \
\ew{\hat X_V^\dag\hat G \hat a^\dag_{V} \hat a^\ndg_{H} \hat a^\ndg_{H} } \right). \nt
\end{align}
Beside its damping due to cavity losses\cite{Carmichael::99} chosen to be $\kappa=10~\mu$eV, the two-photon correlation $\rho_{VH}$ is driven by two higher-order quantities. Both include an exciton-ground state transition under emission of a photon of opposite polarization as the $\xii$ state would allow, e.g. $\hat G^\dg \hat X_H^{\ndg} \hat a^\dg_{V}$ (the complex conjugate of $\hat G^\dg \hat X_V^{\ndg} \hat a^\ndg_{H}$). The transition process takes place under presence of a photon coherence $\hat a^\dag_{V} \hat a^\ndg_{H}$ generated by the previous biexciton-exciton decay, see light-orange boxes in Fig.~\ref{fig:eom_scheme}.  As these terms already include a single-photon coherence and generate a second one leading to a two-photon coherence, they are exactly the terms one would expect to contribute to $\rho_{VH}$. 
\begin{figure}[bt!]
\centering
\includegraphics[clip, width=\columnwidth]{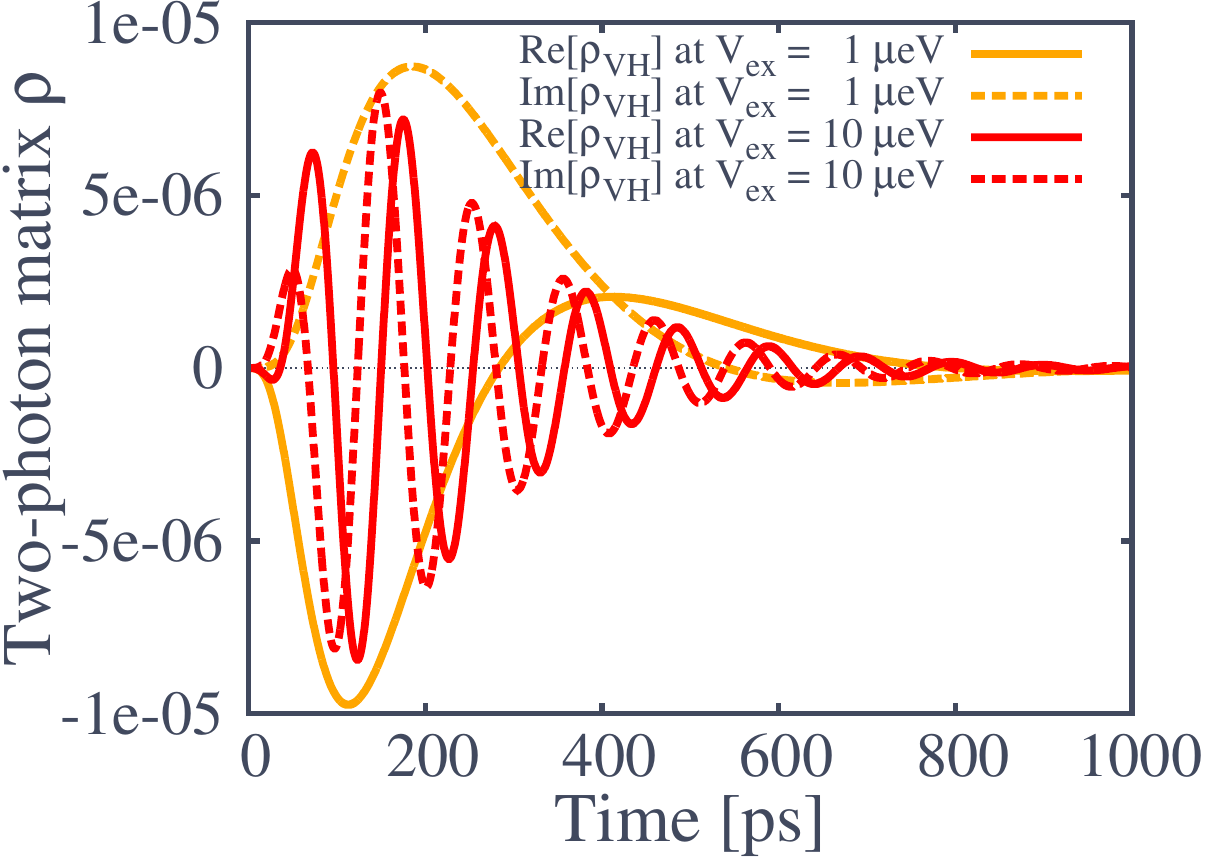}
\caption{Temporal evolution of the non-integrated off-diagonal two-photon density matrix elements at $T=0$~K (real (solid) and imaginary (dashed) part) for $\vex=1 \ \mu$eV (orange) and $\vex=10 \ \mu$eV (red). With increasing $\vex$ the oscillations become more rapid.} 
\label{fig:rhoVHTime}
\end{figure}
For a small FSS, the fixed cavity frequencies $\omega_{V}^{\text{cav}}$ and $\omega_{H}^{\text{cav}} $ are in near-resonance and $\rho_{VH}$ will slowly oscillate on the timescale given by the corresponding FSS, see the orange curves (all at $T=0$~K) in Fig.~\ref{fig:rhoVHTime}. For an increasing FSS on the other hand both frequencies are detuned and $\rho_{VH}$ shows a strong oscillating behavior, compare red curves in Fig.~\ref{fig:rhoVHTime}. Here, the temporal mean value of $\rho_{VH}$ is close to zero and thus no entanglement in a measurement is observed, compare with red curve (all at $T=0$~K) in Fig.~\ref{fig:integratedrho} for the integrated $\rho_{VH}$. In a physical interpretation that means the two different decay paths are distinguishable, so either the photons are entirely emitted in the $H$ or $V$ cascade, but there is no overlap which is only generated by transitions like $\hat G^\dg \hat X^{\ndg}_H \hat a^\dg_{V}$, containing both $V,H$ indices. The which-path information is conserved. If there is an uncertainty in the decay path, the photons become partially polarization-entangled. 

The quantities in this section are damped by the LO-phonon-assisted WL influence as motivated in section \ref{sec:wetting-layer-hamiltonian}. 
The calculated damping rates $\Gamma_{w}$ from Eqs.~\eqref{eq.qdwl.ratengl.pol} and \eqref{eq.qdwl.ratengl.dens} correspond to a $T_1$-time and are incorporated like the radiative dephasing $\Gamma_\text{rad}$ in the Weisskopf-Wigner theory.\cite{Scully::97,Carmele:PhysRevB:09} Both occur and lead to an overall damping of $\Gamma=\Gamma_\text{rad}+\Gamma_{w}$:
\begin{align*}
 \partial _t \left\langle {\hat B^\dag  \hat B} \right\rangle  &\propto  - 4\Gamma \left\langle {\hat B^\dag  \hat B} \right\rangle , \\ 
 \partial _t \left\langle {\hat X_{i} ^\dag  \hat B\hat a^{\ndg}_{i} } \right\rangle &\propto  - 3\Gamma \left\langle {\hat X_{i} ^\dag  \hat B\hat a^{\ndg}_{i} } \right\rangle , \\ 
 \partial_{t} \ew{\hat X^{\dag}_{i}\hat X^{\ndg}_{i}\hat a^{\dag}_{i}\hat a^{\ndg}_{i}} &\propto -2\Gamma \ew{\hat X^{\dag}_{i}\hat X^{\ndg}_{i}\hat a^{\dag}_{i}\hat a^{\ndg}_{i}},\\
 \partial _t \left\langle {\hat G^\dag  \hat X^{\ndg}_{i} \hat a^{\ndg}_{i} } \right\rangle  &\propto  - 1\Gamma \left\langle {\hat G^\dag  \hat X^{\ndg}_{i} \hat a^{\ndg}_{i} } \right\rangle .
 \end{align*}
The exciton $\hat X_{i}$ is damped by the mere presence of the empty  WL states as they disturb the system and lead to a decoherence. This introduces a temperature dependence to the cascade, which is inherit to the perturbation induced by the coupling to the WL and thus included in $\Gamma_w=\Gamma_w(T)$ motivated in Sec.~\ref{sec:wetting-layer-hamiltonian}. 

By numerically solving the set of Eqs. (\ref{eq:eomRhoVH}, \ref{eq:eomDriving1}-\ref{eq:eomLast}) given in App.~\ref{app:eom} we will investigate how the temperature affects the concurrence. Later, the complete temporal dynamics of the cascade presented already partially in this section is resolved.
\section{Results - Dynamics of the biexciton cascade and quality of entanglement}\label{sec:results}

\begin{figure}[bt!]
\centering
\includegraphics[clip, width=\columnwidth]{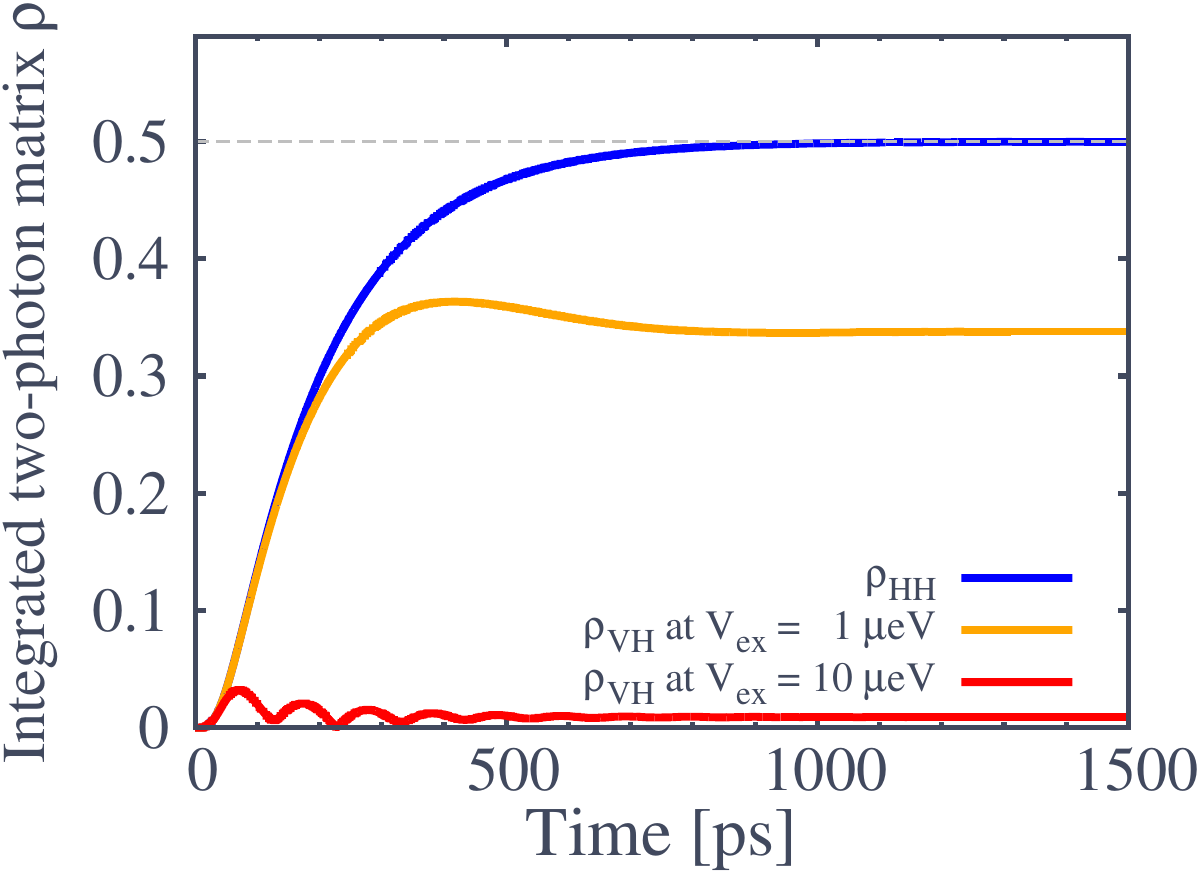}
\caption{Integrated two-photon matrix elements at $T=0$~K. Their steady-state values give the quantum state tomography.} 
\label{fig:integratedrho}
\end{figure}
Since $\rho^\text{pt}$ is experimentally reconstructed by photo-counting experiments all elements of $\rho^\text{pt}$ are given by time-averaging.\cite{Troiani:PhysRevB:06} Recalling and employing Eq.~\eqref{eq:g2} to the results of Fig.  \ref{fig:rhoVHTime}, where already the temporal evolution of $\rho^\text{pt}$ elements is given we get its integrated elements.  As can be seen in Fig.~\ref{fig:integratedrho} the diagonal elements have a continuous positive slope and start to saturate around 0.5~ns. The steady state which determines the quantum state tomography is reached at 1~ns. However, the situation is very different for the off-diagonal elements, as they are complex quantities that oscillate when $V_\text{ex}\neq 0$. Its absolute value (important for the concurrence) shows a non-monotonous behavior in Fig.~\ref{fig:integratedrho}. Obviously, the concurrence is lost for a FSS higher than $V_\text{ex} = 10~\mu$eV. 

\begin{figure}[bt!]
\centering
\includegraphics[clip, width=\columnwidth]{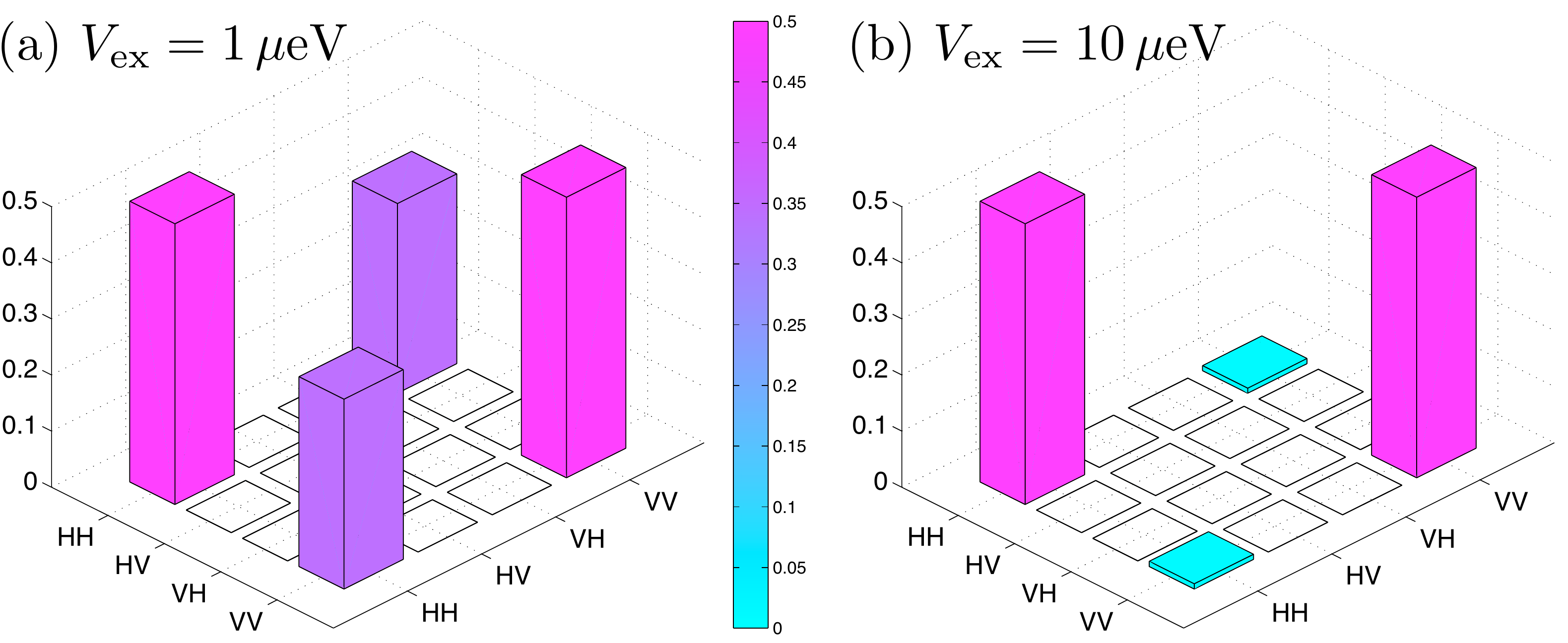}
\caption{Quantum state tomography for (a) $\vex=1~\mu$eV and (b) $\vex=10~\mu$eV. } 
\label{fig:qst0}
\end{figure}

\begin{figure}[bt!]
\centering
\includegraphics[clip, width=\columnwidth]{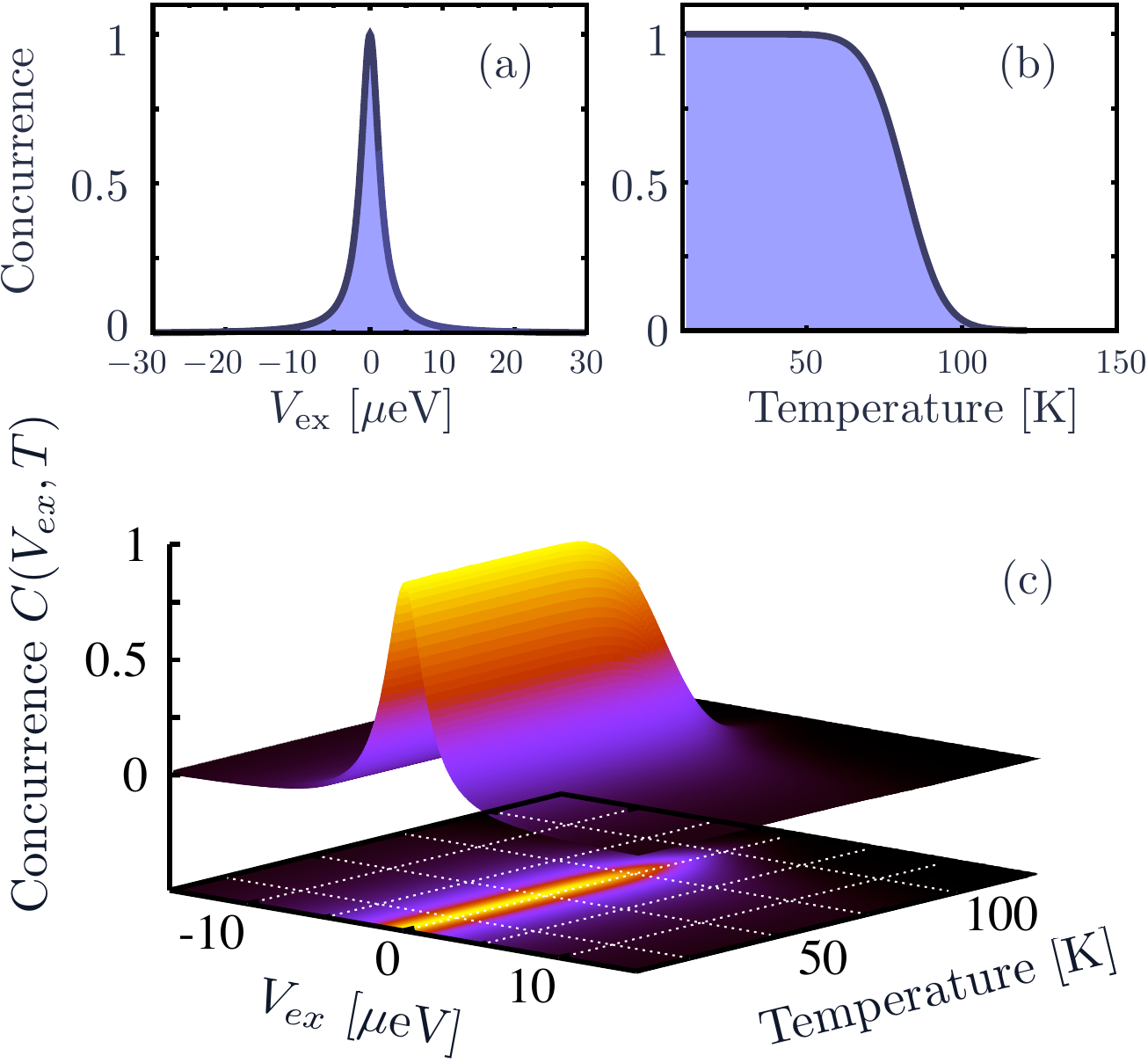}
\caption{Plot of the concurrence. (a) displays $C$ at 0~K as a function of $\vex$ only. (b) illustrates the influence of the temperature due to phonon-induced WL influence at $\vex = 0~\mu$eV. Both temperature and $\vex$ dependence are shown in (c).} \label{fig:both_dependences}
\label{fig:3dTemp}
\end{figure}

Since all steady-state elements of $\rho^\text{pt}$ are given by time-averaging, $C$ will drop for increasing $\vex$ as the integrated $\rho_{VH}$ did. This can be clearly seen in the quantum state tomography shown in Fig.~\ref{fig:qst0}. The diagonal and off-diagonal contributions are still in the same order of magnitude for $\vex= 1~\mu$eV cp. Fig.~\ref{fig:qst0}(a), but a loss can already be seen. For a larger splitting $\rho_{VH}$ vanishes in Fig.~\ref{fig:qst0}(b). Figure~\ref{fig:3dTemp} constitutes the central result of this work -- a temperature dependent study of entanglement. First, Fig.~\ref{fig:3dTemp}(a) shows how the entanglement is lost with increasing FFS as a continuous function of $\vex$. Here, the FWHM is determined by the values of the Coulomb parameters.
When temperature effects of the WL states are taken into account, the concurrence can be spoiled even in the ideal situation of degenerate exciton states, see Fig.~\ref{fig:3dTemp}(b). For low temperatures $C$ will remain unaffected by the WL-induced dephasing $\Gamma$, since the scattering times are well above 1 ns, cp. Fig.~\ref{fig:mrd_rated}. Starting at approximately 80~K the WL starts to affect $C$ as $\Gamma$ reaches 1~ns$^{-1}$, which corresponds to an energy of $0.7~\mu$eV close to the optical coupling strength of $M=1~\mu$eV.  The entanglement decreases for zero $\vex$ until it is entirely lost for temperatures beyond 100~K. For a higher FSS with $\vex\ne0$, Fig.~\ref{fig:3dTemp}(c) shows, that the degree of entanglement is lost slightly earlier around 80~K.  

\begin{figure}[bt!]
\centering
\includegraphics[clip, width=\columnwidth]{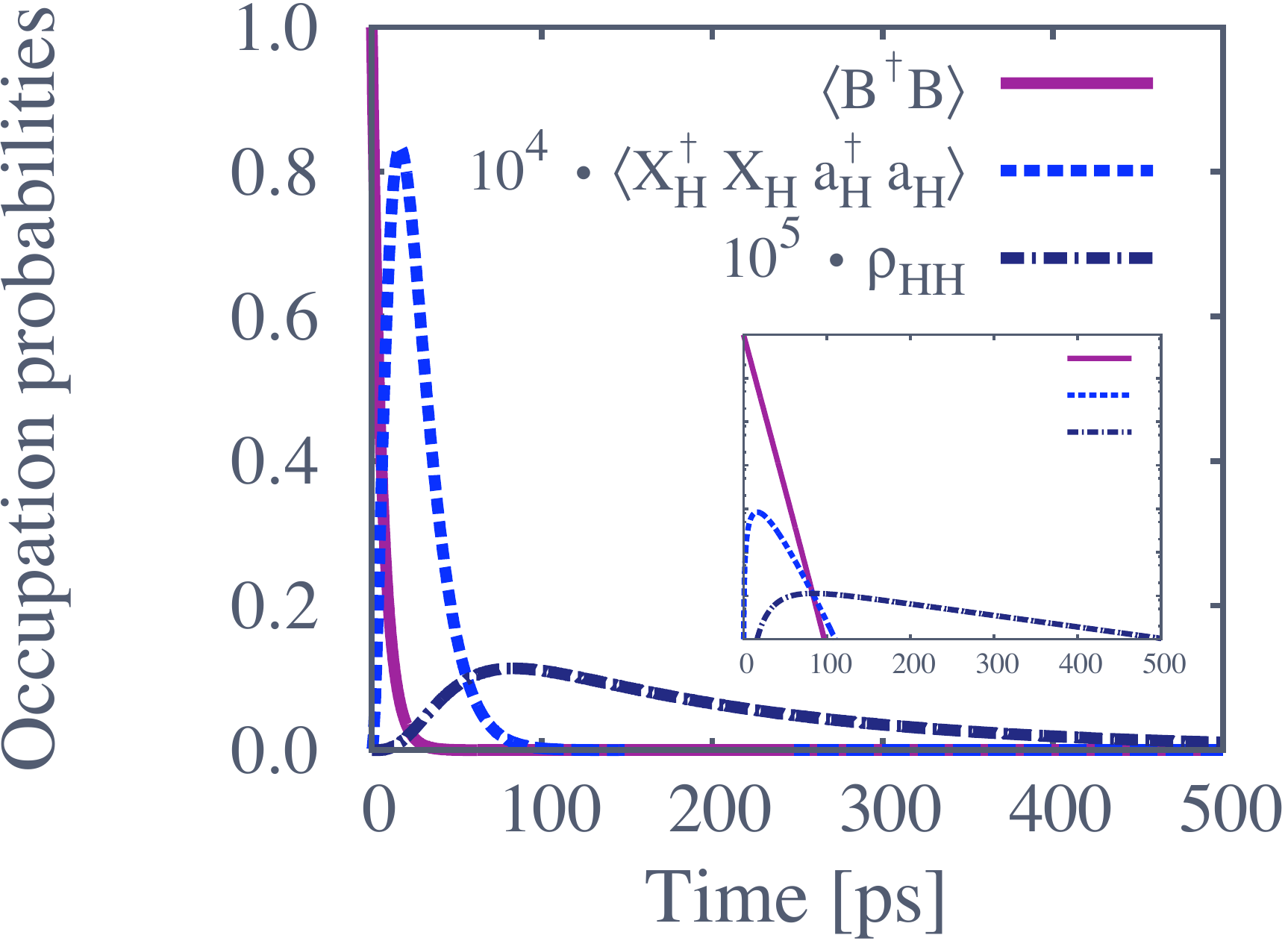}\caption{Dynamics of the cascade biexciton $\rightarrow$ exciton-and-one-photon $\rightarrow$ ground-state-and-two-photons. Results for vanishing $V_{\text{ex}}=0~\mu$eV at $T=0$~K. $\ew{\hat X^\dag_H \hat X_H \hat a^\dag_{H}\hat a_{H}}$ and  $\rho_{HH}$ are enlarged by a factor of $10^{4}$ and $10^{5}$, respectively. The inset shows the dynamics on a  logarithmic scale.} 
\label{fig:timeCascadeFSS0}
\end{figure}

Finally, to pick up on the topic of temporal dynamics of the cascade already addressed in Sec.~\ref{subsec:eom} let us consider a direct, single path leading to no entanglement. Therefore, we will follow the blue $HH$ (left) path in Fig.~\ref{fig:eom_scheme}. The biexciton density $\langle\hat B^\dag \hat B\rangle$ decays exponentially with 4 $\Gamma$ giving rise to an intermediate coupled exciton-photon state $\ew{\hat X^\dag_H \hat X^\ndg_H \hat a^\dag_{H} \hat a^\ndg_{H}}$. Subsequently, when this state is sufficiently populated it decays under emission of the exciton-photon and a two-photon density $\rho_{HH}$ builds up. In the given range of parameters (see table~\ref{tab:param}), Fig.~\ref{fig:timeCascadeFSS0} shows that the decay cascade happens on a ps time scale. Even at low temperatures and $V_\text{ex}=0~\mu$eV, due to high cavity loss $\kappa$ (compared to the optical coupling strength $M$) both $\ew{\hat X^\dag_H \hat X^\ndg_{H} \hat a^\dag_{H}\hat a^\ndg_{H}}$ and $\rho_{HH}$ are only weakly occupied. The inset in Fig.~\ref{fig:timeCascadeFSS0} is a logarithmic plot of the dynamics which clearly shows the different lifetimes of the involved quantities.
\section{Conclusion}
In summary, we showed that -- based on a Heisenberg-equation approach -- the density matrix of the photon-polarization subspace of a biexciton, all intermediate occurring states of the cascade as well as their dynamics can be microscopically calculated. The interaction of the dot states with the WL via LO-phonons was included within this approach and gave rise to a strong reduction of the concurrence for temperatures above 100~K for typical InGaAs self-assembled QDs. 

The diagonal interaction of the QD states with longitudinal phonons is a major contribution to dephasing effects,\cite{Borri:PhysRevLett:01,Milde:PhysRevB:08} which will ultimately influence the quality of entanglement.\cite{Hohenester:PhysRevLett:07} 

Our conclusion is, that regardless of their impact, the inherit coupling to the WL imposes another fundamental limit to high temperature generation of polarisation-entangled photons in solid state devices.
\begin{acknowledgments}
We would like to thank Stephen Hughes for helpful discussions. This work was financially supported by the Deutsche Forschungsgemeinschaft within the Sonderforschungsbereich 787 ``Nanophotonik''.\end{acknowledgments}
\appendix\label{sec:appendix}
\section{Carrier-carrier interaction and exciton representation}\label{app:ccHam}
 When constructing the single-particle basis which is to be diagonalized we can employ the fact that only a fraction of all possible QD states will contribute. First, not all transitions are optically active. A transition of a conduction band electron under emission of a photon must conserve the angular momentum. The electron-photon matrix element leads to selection rules that have to be obeyed and so only electron-hole pairs with opposite spins couple. Second, and more important, the Pauli-Principle forbids two carriers to be in the same state (that is the spin state $s=\ket{J,m_j}$).

With these considerations, only four states remain to determine the system dynamics, defined by the following operators:
\bqn
\hat G &=  \hat h^{\ndg}_{\da} \hat h^{\ndg}_{\ua}\hat h^{\dg}_{\ua} \hat h^{\dg}_{\da} &\quad \text{Ground state operator}, \\
\hat B &= \hat h^{\ndg}_{\ua} \hat h^{\ndg}_{\da} \hat e^{\ndg}_{\ua} \hat e^{\ndg}_{\da} &\quad \text{Biexciton state operator}, \\
\hat X^{ \ndg}_+  &= \hat h^{\ndg}_{\ua} \hat h^{\ndg}_{\da} \hat h^{\dg}_{\da} \hat e^{\ndg}_{\da} &\quad \text{Exciton state operator 1}, \label{eq:ex1}\\
\hat X^{ \ndg}_{{-}} &= \hat h^{\ndg}_{\da} \hat h^{\ndg}_{\ua} \hat h^{\dg}_{\ua} \hat e^{\ndg}_{\ua} &\quad \text{Exciton state operator 2}\label{eq:ex2}.
\eqn
Commuting these operators with $\hat H_\text{Coul}=\hat H_{\text{QD},0}^{\text{c}} +\hat  H^{\text{c}}_\text{c-c}$, we note, that ground state $\hat G$ and biexciton operator $\hat B$ are already diagonal:
\begin{align}
\left[ \hat H_\text{Coul} , \hat G \right] &= \hbar \omega_G \ \hat G, & \left[ \hat H_\text{Coul} , \hat B \right] &= \hbar \omega_B \ \hat  B ,
\end{align}
with $\hbar\omega_G = 0$, choosing zero as ground state energy, and $\hbar\omega_B = 4E_C + 2\hbar\Omega_0 + V^\text{ex}_{\downarrow\downarrow} +V^\text{ex}_{\uparrow\uparrow}$ with $E_C =\frac{1}{2} \left( V^{cc}+V^{vv}-2V^{vc} \right)$. Investigating the exciton operators $\hat X_\pm$, we find:
\begin{align}
\left[ \hat H_\text{Coul} , \hat X_+ \right] &= (E_C+\hbar\Omega_0 + V^\text{ex}_{\downarrow\downarrow}) \hat X_+ -V^\text{ex}_{\downarrow\uparrow} \ \hat X_- \\
\left[ \hat H_\text{Coul} , \hat X_- \right] &= (E_C+\hbar\Omega_0 + V^\text{ex}_{\uparrow\uparrow}) \hat X_- - V^\text{ex}_{\uparrow\downarrow} \ \hat X_+.
\end{align}
The exciton operators in Eqs.~\eqref{eq:ex1} and~\eqref{eq:ex2} emit into circular polarized light modes ($\sigma_{+}$ and $\sigma_{-}$), if the non-diagonal element $V^\text{ex}_{\uparrow\downarrow}$ is zero. In a reduced $C_{\text{2v}}$ symmetry of strained dots, the off-diagonal element $V^\text{ex}_{\uparrow\downarrow}$ is non-zero and leads to a superposition of the exciton states, which are not eigenstates of $\hat H_\text{Coul}$.\cite{Danckwerts:PhysRevB:06} 
For convenience, via solving a diagonalization problem, new exciton operators $\hat X_{H/V}$ are introduced. Here, $H$ and $V$ refer to the linear polarization of the emitted photons in the biexciton cascade: 
\begin{align}
\hat X^{\ndg}_H =&  u^H_1 \hat X^{ \ndg}_++u^{H}_{2}\hat X^{ \ndg}_-,&
\hat X^{\ndg}_V =&  u^V_1 \hat X^{ \ndg}_++u^{V}_{2}\hat X^{ \ndg}_-.
\end{align}
The unitary transformation coefficients are given by 
\begin{align*}
u_1^H&=-u_2^V=-(1+\Delta_e^2)^{-\frac{1}{2}}\Delta_e,\nt \\
u_2^H&=u_1^V=(1+\Delta_e^2)^{-\frac{1}{2}},
\end{align*}
where $\Delta_e=V_\text{ex}(E_C+\hbar\Omega_0 + V^\text{ex}_{\downarrow\downarrow}-\hbar\omega_H)^{-1}=-V_\text{ex}(E_C+\hbar\Omega_0 + V^\text{ex}_{\ua\ua}-\hbar\omega_V)^{-1}$ and $V_\text{ex}=V^\text{ex}_{\da\ua}=V^\text{ex}_{\ua\da}$. 
Within the new excitonic basis the electron operators ($\hat G^{(\dg)}$, $\hat X^{(\dg)}_H$, $\hat X^{(\dg)}_V$, $\hat B^{(\dg)}$) are eigenvectors to $\hat H_{\text{QD},0}^{\text{c}} +\hat  H^{\text{c-c}}_\text{QD}$. Their corresponding eigenvalues are ($\hbar\omega_G, \hbar\omega_H, \hbar\omega_V, \hbar\omega_B$) with exciton energies $\hbar\omega_H$ and $\hbar\omega_V$ in the two-particle basis:
\bqn
\hbar \omega_{H/V} = 
\frac{1}{2}
\left(
2 E_C + 2 \hbar\Omega_0 + 
V^\text{ex}_{\downarrow\downarrow} + V^\text{ex}_{\ua\ua} \pm
\delta
\right) .
\eqn
Responsible for the fine structure splitting $\delta$, compare Fig.~\ref{fig:non_local_cascade_scheme}, is the exchange splitting $V^\text{ex}_{\ua\da}$, which describes the repulsion and attraction forces induced by different spin-conformations of electrons and holes. In the most general case, the exciton states could differ in energy due to contributions like $V^\text{ex}_{\ua\ua}$ and $V^\text{ex}_{\da\da}$. Given these elements, the most general fine structure splitting can now be expressed quantitatively as
$\delta := \sqrt{ \left( V^\text{ex}_{\ua\ua} - V^\text{ex}_{\da\da} \right)^2 + 4 |V^\text{ex}_{\ua\da}|^2  } $.
However, it is reasonable to assume that in semiconductor QD no spin-preferences exist, thus $V^\text{ex}_{\ua\ua} - V^\text{ex}_{\da\da}=0$, which leads to 
\bqn
\delta=2 \left|V^\text{ex}_{\ua\da}\right|.
\eqn 
In our case of no spin-preferences, where $(V^{ex}_{\downarrow\downarrow}-V^{ex}_{\ua\ua})=0$, it follows that $\Delta_e = -1$, $u_1^H=u_2^H=u_1^V=-u_2^V=\frac{1}{\sqrt{2}}$,
and thus explicitly 
\begin{align}
\hat X^{\ndg}_H =&  \frac{1}{\sqrt{2}}\left( \hat X^{ \ndg}_++\hat X^{ \ndg}_-\right),&
\hat X^{\ndg}_V =& \frac{1}{\sqrt{2}}\left( \hat X^{ \ndg}_+-\hat X^{ \ndg}_-\right).
\end{align}
\section{QD electron-photon interaction}\label{app:eptHam}
Starting with Eq.~\eqref{eq:Helpt}, we now switch to the new exciton operators $\hat X_{H/V}$ by inserting the unity relation of the electron-hole picture into Eq.~\eqref{eq:Helpt}. After normal ordering and using the two-electron assumption, the electron-light interaction can be written as:\cite{Carmele:PhysRevB:09}
\begin{eqnarray}
\hat H_{\text{QD}}^\text{c-pt} = &\phantom+& \hbar M \sum_{i}
\left(
\hat G^\dg \hat X_{+} 
+
\hat X_{-}\hat B
\right)
\hat a^\dg_{i\sigma_+} \\
&+& \hbar M \sum_{i}
\left(
\hat G^\dg \hat X_{-}
+
\hat X_{+} \hat B 
\right) 
\hat a^\dg_{i\sigma_-}
 + \text{h.a.} .\nt
\end{eqnarray}
The electron-light interaction Hamiltonian $\hat H_{\text{QD}}^\text{c-pt}$ is transformed, when the exciton operators are replaced with:
\begin{align}
\hat X^{\ndg}_{+} =&  \frac{1}{\sqrt{2}}\left( \hat X_H+\hat X_V\right),&
\hat X^{\ndg}_{-} =& \frac{1}{\sqrt{2}}\left( \hat X_H-\hat X_V\right).
\end{align}
It is convenient to define new photon operators:\begin{align}
\hat a^\dg_{iH} &= \frac{1}{\sqrt{2}} 
\left(
\hat a^\dg_{i\sigma_+}
+
\hat a^\dg_{i\sigma_-}
\right),\nt\\
\hat a^\dg_{iV} &= \frac{1}{\sqrt{2}} 
\left(
\hat a^\dg_{i\sigma_+}
-
\hat a^\dg_{i\sigma_-}
\right).
\end{align}
The Hamiltonian now takes the form:
\begin{eqnarray}
\hat H_{\text{QD}}^\text{c-pt} = &\phantom+& \hbar M \sum_{i}
\left(
\hat G^\dg \hat X_H^{\ndg} \ \hat a^\dg_{iH} 
+
\hat G^\dg \hat X_V^{\ndg}  \ \hat a^\dg_{iV} 
\right) \\
&+& \hbar M \sum_{i}
\left(
\hat X_H^{\dg} \hat B \ \hat a^\dg_{iH}
-
\hat X_V^{\dg} \hat B \ \hat a^\dg_{iV}
\right) 
 + \text{h.a.} .\nt
\end{eqnarray}
\section{Equations of motion}\label{app:eom}
The temporal evolution of the driving terms in Eq.~\eqref{eq:eomRhoVH} is given by
\begin{align}
\partial_t &\ew{\hat G^\dg \hat X^{\phdg}_H \hat a^\dag_{V} \hat a^\dag_{V} \hat a^\ndg_{H}}
=\nt\\
 \phantom{+}&\ii\left( 
       -\omega_H + 2 \omega_V^\text{cav} - \omega_H^\text{cav} 
       +\ii \Gamma + 3\ii\kappa  
     \right) 
     \ew{\hat G^\dg \hat X^{\phdg}_H \hat a^\dag_{V} \hat a^\dag_{V} \hat a^\ndg_{H}} \nt\\
-&2\ii M \ew{\hat X_V^\dag \hat X^{\phdg}_H \hat a^\dag_{V} \hat a^\ndg_{H}} 
+\ii M\ew{\hat G^\dg \hat B \hat a^\dag_{V} \hat a^\dag_{V} }\label{eq:eomDriving1}
\end{align}
and
\begin{align}
\partial_t &\ew{\hat X_V^\dag \hat G\ \hat a^\dag_{V} \hat a^\ndg_{H} \hat a^\ndg_{H}}
=\nt\\
\phantom{+}&  \ii\left( 
         \omega_V + \omega_V^\text{cav} -2 \omega_H^\text{cav} 
       +\ii \Gamma +3\ii\kappa  
     \right) 
     \ew{\hat X_V^\dag \hat G\ \hat a^\dag_{V} \hat a^\ndg_{H} \hat a^\ndg_{H}}\nt\\
+ &2\ii M 
\ew{\hat X_V^\dag \hat X^{\phdg}_H \hat a^\dag_{V} \hat a^\ndg_{H} }
+\ii M  \ew{\hat B^\dag \hat G\ \hat a^\ndg_{H} \hat a^\ndg_{H}}.  \label{eq:eomDriving2}
\end{align}
The driving terms of the two-photon density matrix in turn couple to combined exciton- and photon coherences $\hat X_V^\dag \hat X_H^{\ndg} \hat a^\dag_{V} \hat a^\ndg_{H}$ and to the direct decay channel from $\bb$ to $\GG$ emitting two photons with the same polarization $\hat G^\dg \hat B \hat a^\dag_{V} \hat a^\dag_{V}$., see orange box in Fig.~\ref{fig:eom_scheme}. Crucial for entangling the two decay paths is the exciton coherence, assisted by a photon coherence, see red box in Fig.~\ref{fig:eom_scheme}:
\begin{align}
\partial_t& \ew{\hat X_V^\dag \hat X^{\phdg}_H \hat a^\dag_{V} \hat a^\ndg_{H} } 
=\nt\\
\phantom{+}& \ii\left( 
         \omega_V - \omega_H + \omega_V^\text{cav} - \omega_H^\text{cav} 
       +2\ii \Gamma +2\ii\kappa  
     \right)      \ew{\hat X_V^\dag \hat X^{\phdg}_H \hat a^\dag_{V} \hat a^\ndg_{H} }\nt\\
\nonumber
-&
\ii M     \ \ew{\hat G^\dg \hat X^{\phdg}_H \hat a^\dag_{V} \hat a^\dag_{V} \hat a^\ndg_{H} }
+
\ii M \ \ew{\hat B^\dag \hat X^{\phdg}_H \hat a^\ndg_{H} } \\
+&\ii M   \ew{\hat X_V^\dag \hat G\ \hat a^\dag_{V} \hat a^\ndg_{H} \hat a^\ndg_{H} }
 + \ii M \ew{\hat X_V^\dag \hat B \hat a^\dag_{V} } .\label{eq:exptcoh}
\end{align}
In this equation, the two paths interfere. The influence in the two-particle correlation $\ew{\hat X_V^\dag \hat X^{\phdg}_H \hat a^\dag_{V} \hat a^\ndg_{H} }$ increases the degree of entanglement as this term couples back to the driving terms of $\rho_{VH}$, Eq.~\eqref{eq:eomDriving1} and~\eqref{eq:eomDriving2}.  Here again the resonance condition of the frequencies is essential ($\omega_V - \omega_H = \omega^\text{cav}_V - \omega^\text{cav}_H = \delta$):  A high detuning $\delta$ will diminish the contribution of Eq.~\eqref{eq:exptcoh} to the cascade and both paths cannot interfere.

The other characteristic and important quantity in the two-electron biexciton-cascade situation (cp. with two coupled QDs \cite{Carmele:PhysRevB:09}) are the two-photon polarizations
\begin{align}
\partial_t& \ew{\hat G^\dg \hat B \hat a^\dag_{V} \hat a^\dag_{V} }
=\nt \\
\phantom+&\ii \left( -\omega_B + 2 \omega_V^\text{cav}
       +2\ii \Gamma + 2\ii\kappa 
     \right) 
     \ew{\hat G^\dg \hat B \hat a^\dag_{V} \hat a^\dag_{V} }\label{eq:eomGBaa} \\
\nonumber
+& 
\ii M \ \ew{\hat G^\dg \hat X^{\phdg}_H \hat a^\dag_{V} \hat a^\dag_{V} \hat a^\phdg_{H} }
-  
\ii M \ \ew{\hat G^\dg \hat X_V^\phdg \hat a^\dag_{V} \hat a^\dag_{V} \hat a^\phdg_{V} } \\
\nonumber
-&
2\ii M \ \ew{\hat X_V^\dag \hat B \hat a^\dag_{V} }
\end{align}
and
\begin{align}
\nonumber
\partial_t & \ew{\hat B^\dag \hat G\ \hat a^\ndg_{H} \hat a^\ndg_{H} } 
= \nt \\
\phantom-&\ii \left( \omega_B - 2 \omega_H^\text{cav}
       +2\ii \Gamma + 2\ii\kappa 
     \right)  
     \ew{\hat B^\dag \hat G\ \hat a^\ndg_{H} \hat a^\ndg_{H} } \\
\nonumber
-& 
\ii M \ \ew{\hat X_H^\dag \hat G \ \hat a^\dg_{H} \hat a^\ndg_{H} \hat a^\ndg_{H} } 
+
\ii M \ \ew{\hat X_V^\dag \hat G \ \hat a^\dag_{V} \hat a^\ndg_{H} \hat a^\ndg_{H} } \\
+& 
2\ii M \ \ew{\hat B^\dag \hat X^{\ndg}_H \hat a^\ndg_{H} }.
\end{align}
Each path in the cascade has one biexciton-to-ground  state transition like $\hat G^\dg \hat B \hat a^\dag_{V} \hat a^\dag_{V}$. Its dynamics couples the biexciton-to-exciton transition $\hat X_V^\dag \hat B \hat a^{\dg}_{V}$ with both exciton-to-ground state transitions $\hat G^\dg \hat X_{i}$. Remarkably, the origin of the entanglement is directly visible, since a quantity of a different path enters in Eq.~\eqref{eq:eomGBaa}: $\ew{\hat G^\dg \hat X^{\ndg}_H \hat a^\dag_{V} \hat a^\dag_{V} \hat a^\ndg_{H}}$. Here again, the two paths interfere. For maximum entanglement the contributions of the different paths $\hat G^\dg \hat X_{H}$ and $\hat G^\dg \hat X^{\ndg}_{V}$ to the expectation values should be equally weighted. The photon-assisted biexciton-to-exciton transition enters in the two-photon polarization and drives this quantity via the biexciton decay:
\begin{align}
\partial_t& \ew{\hat B^\dag \hat X^\ndg_H \hat a^\ndg_{H} }
=\nt\\
\phantom-&\ii 
     \left( 
        \omega_B - \omega_H - \omega_H^\text{cav}
       +3\ii \Gamma + \ii\kappa 
     \right)  
     \ew{\hat B^\dag \hat X^\ndg_H \hat a^\ndg_{H} } \\
\nonumber
-&
\ii M \ \ew{\hat X_H^\dag \hat X^\ndg_H \hat a^\dg_{H} \hat a^\ndg_{H} }
+
\ii M \ \ew{\hat X_V^\dag \hat X^\ndg_H \hat a^\dag_{V} \hat a^\ndg_{H}} \\
\nonumber
+&
\ii M \ew{\hat B^\dag \hat B} 
+
\ii M \ \ew{ \hat B^\dag \hat G \ \hat a^\ndg_{H} \hat a^\ndg_{H} } ,
\end{align}
\begin{align}
\partial_t & \ew{\hat X_V^\dag \hat B \hat a^\dag_{V} }
=\nt\\
\phantom-&\ii 
     \left( 
        -\omega_B + \omega_V + \omega_V^\text{cav}
       +3\ii \Gamma +\ii\kappa 
     \right) 
     \ew{\hat X_V^\dag \hat B \hat a^\dag_{V} } \\
\nonumber
-&
\ii M     \ \ew{\hat G^\dg \hat B \hat a^\dag_{V} \hat a^\dag_{V} } 
+
\ii M \ \ew{\hat B^\dag\hat B} \\
\nonumber
+&
\ii M \ \ew{\hat X_V^\dag \hat X^{\ndg}_H \hat a^\dag_{V} \hat a^\ndg_{H} }
-
\ii M \ \ew{\hat X_V^\dag \hat X_V^\ndg \hat a^\dag_{V} \hat a^\ndg_{V}  }.
\end{align}
The occurring biexciton as well as the intermediate exciton-photon densities are driven by the biexciton-exciton transition $\ew{\hat X_i^\dag \hat B \hat a^\dag_{i} }$: 
\begin{align}
\partial_t & \ew{\hat X_H^\dag \hat X_H^{\ndg} \hat a^\dg_{H} \hat a^\ndg_{H} }
= \nt\\
-& \left( 
       2 \Gamma + 2\kappa 
     \right) 
  \ew{\hat X_H^\dag \hat X_H^{\ndg} \hat a^\dg_{H} \hat a^\ndg_{H} }  \\
\nonumber
-&
2 \ \im\left( M\ew{\hat X_H^\dag \hat B \hat a^\dg_{H} }
+  
M \ew{\hat X_H^\dag \hat G \ \hat a^\dg_{H} \hat a^\ndg_{H} \hat a^\ndg_{H} }\right),
\end{align}
\begin{align}
\partial_t & \ew{\hat X_V^\dag \hat X_V^\dag \hat a^\dag_{V} \hat a^\ndg_{V} }
= \nt\\
-& \left( 
       2 \Gamma + 2\kappa 
     \right) 
  \ew{\hat X_V^\dag \hat X_V^\dag \hat a^\dag_{V} \hat a^\ndg_{V} } \\
\nonumber
+&
2\ \im  \left( M\ew{\hat X_V^\dag \hat B \hat a^\dag_{V} }
-
 M\ew{\hat X_V^\dag \hat G \ \hat a^\dag_{V} \hat a^\ndg_{V} \hat a^\ndg_{V} }\right).
\end{align}
From the perspective of the cascade our course of action so far put the cart before the horse since the actual dynamics start with a loaded biexciton density $\ew{\hat B^\dag \hat B}$. In the visualization of the complex interplay, Fig.~\ref{fig:eom_scheme}, we followed a bottom-to-top trail through the cascade, starting with the concurrence determining $\rho_{V H}$. The biexciton $\ew{\hat B^\dag \hat B}$ as the top element of the scheme decays via the $H$ or the $V$ intermediate exciton-to-ground-state path 
\begin{align}
\partial_t& \ew{\hat B^\dag \hat B}
= 
- 4 \Gamma 
        \ew{\hat B^\dag \hat B} \\
\nonumber
+& 
2 \ \im \left(M \ew{\hat X_H^\dag \hat B \hat a^\dg_{H} }
-
 M \ew{\hat X_V^\dag \hat B \hat a^\dag_{V} }\right).
\end{align}
To complete the set of equation, two higher-order photon-assisted exciton-to-ground state transitions of the direct and thus not entangled path are necessary:
\begin{align}
\partial_t &\ew{\hat G^\dg \hat X^{\ndg}_H \hat a^\dg_{H} \hat a^\dg_{H} \hat a^\ndg_{H}}
=\nt\\
\phantom-&\ii
     \left( -\omega_H + \omega_H^\text{cav}
      +\ii\Gamma + 3\ii\kappa 
     \right)  
     \ew{\hat G^\dg \hat X^{\ndg}_H \hat a^\dg_{H} \hat a^\dg_{H} \hat a^\ndg_{H}}\label{eq:gxhaaa} \\
\nonumber
-&
2\ii M \  \ew{\hat X_H^\dag \hat X^{\ndg}_H \hat a^\dg_{H} \hat a^\ndg_{H}} 
+
\ii M\ \ew{\hat G^\dg \hat B \hat a^\dg_{H} \hat a^\dg_{H} },
\end{align}
\begin{align}
\partial_t &\ew{\hat G^\dg \hat X_V^\ndg\hat a^\dag_{V} \hat a^\dag_{V} \hat a^\ndg_{V}}
=\nt\\
\phantom-&\ii
     \left( -\omega_V + \omega_V^\text{cav}
      + \ii\Gamma + 3\ii\kappa 
     \right) 
     \ew{\hat G^\dg \hat X_V^\ndg\hat a^\dag_{V} \hat a^\dag_{V} \hat a^\ndg_{V}} \label{eq:gxvaaa} \\
\nonumber
-&
2\ii M \  \ew{\hat X_V^\dag \hat X_V^\ndg \hat a^\dag_{V} \hat a^\ndg_{V}} 
-
\ii M\ \ew{\hat G^\dg \hat B \hat a^\dag_{V} \hat a^\dag_{V} }.
\end{align}
With these polarization Eq. (\ref{eq:gxhaaa}-\ref{eq:gxvaaa}), the diagonal elements $i=H,V$ of the density matrix of the polarization subspace are given, too:
\begin{align}
\partial_t & 
\ew{\hat a^\dg_{i} \hat a^\dg_{i} \hat a^\ndg_{i} \hat a^\ndg_{i}}
=
\nt\\
-&4\kappa
 \ \ew{\hat a^\dg_{i} \hat a^\dg_{i} \hat a^\ndg_{i} \hat a^\ndg_{i}}
- 
4 \ \im 
\left(
 M \ew{\hat G^\dg \hat X^\dg_i \hat a^\dg_{i} \hat a^\dg_{i} \hat a^\ndg_i}
\right)\label{eq:eomLast}.
\end{align}

\end{document}